\documentclass[fleqn]{elsarticle}
\biboptions{sort&compress}

\usepackage{amsmath, amssymb, bm} 
\usepackage[utf8]{inputenc} 
\usepackage{float}
\usepackage{indentfirst} 
\usepackage{xcolor} 
\usepackage{setspace} 
\usepackage{booktabs, array} 
\usepackage{multirow} 
\usepackage{mathtools}
\usepackage{lineno,hyperref}
\usepackage{amsmath}
\usepackage{xcolor}
\usepackage[a4paper, left=1in, right=1in, top=1in, bottom=1in]{geometry}

\journal{arXiv}

\makeatletter
\providecommand{\@corref}[1]{} 
\providecommand{\cnotenum}[1]{} 
\makeatother

\begin{document}

\begin{frontmatter}

\title{Self-Refining Topology Optimization via an LLM-Based Multi-Agent Framework}

\author[unist]{Hyunjee~Park}
\author[unist]{Hayoung~Chung\corref{mycorrespondingauthor}}
\cortext[mycorrespondingauthor]{Corresponding author}
\ead{hychung@unist.ac.kr}

\address[unist]{Department of Mechanical Engineering, Ulsan National Institute of Science and Technology, 50 UNIST-gil, Ulju-gun, Ulsan, 44919, Republic of Korea}

\begin{abstract}
Topology optimization is a widely used design method that produces optimized material distributions for prescribed objectives and constraints through well-established numerical algorithms. Throughout the workflow, engineers make a series of decisions ranging from setting and adjusting numerical parameters to assessing whether the converged design meets considerations beyond those explicitly included in the optimization problem, such as physical feasibility. These decisions, which draw on domain expertise, interfere with the autonomous design process.
To address this difficulty, this study presents TopOptAgents, a multi-agent system for automating not only the design process but also decision-making during the key stages of the topology optimization process. 
TopOptAgents consists of six LLM-based agents collaborating through iterative self-refinement cycles spanning problem formulation, validation, code generation and execution, and quality assessment of the optimized structure. This process enables error correction and progressive improvement of both the optimization setup and resulting design. The framework is demonstrated on optimization problems selected to cover a range of settings that differ in their literature coverage and numerical characteristics
The benefits of iterative self-refinement are found to be particularly pronounced for problem classes where the pretrained language model has limited prior exposure, such as formulations whose literature and open-source implementations are comparatively sparse. In such cases, the proposed framework reliably produces converged designs where a single state-of-the-art LLM struggles, suggesting that self-refinement broadens the range of topology optimization problems that LLM-based automation can reliably address.
\end{abstract}


\end{frontmatter}

\section{Introduction}
\label{sec:Intro}

Topology optimization is a computational design approach for determining the optimized material distribution within a prescribed design domain subject to certain objectives and constraints \cite{deaton2014survey}. This process involves several numerical procedures, such as finite element analysis that is performed repeatedly upon any change in the topology of material distributions, that in turn changes the sensitivities of the objective and constraints with respect to the design variables, leading to the update of the material distribution \cite{sigmund2013topology}. Thanks to the well-defined methods and techniques involved in the method, topology optimization method has been widely adopted in engineering applications where structural integrity and efficient material distribution are of central importance, including aerospace \cite{zhu2016topology} and automotive engineering \cite{yang1995automotive}. Its application has also expanded beyond classical structural mechanics to areas such as crashworthiness \cite{patel2009crashworthiness}, fluid flow \cite{lundgaard2018revisiting}, and heat transfer \cite{pietropaoli2019three}. Topology optimization has thus become an established computational approach in structural and multidisciplinary design.

Although the method has potential for autonomously conceiving mathematically optimized designs, the optimization process requires engineers' intervention at each step \cite{lynch2019machine}. The standard workflow operates as an iterative cycle of formulation, implementation, and assessment. Engineers must manually translate design requirements into mathematical specifications, adjust numerical parameters, and resolve computational issues \cite{zhang2018finding, ha2024automatic}. Furthermore, evaluating a converged design requires multimodal inspection; engineers analyze textual logs for numerical stability and visually inspect the generated layout for physical feasibility, such as checking for checkerboard patterns or disconnected members \cite{zhou2001checkerboard}. Whenever a result fails to meet these criteria, the engineer must diagnose the root cause, whether in the numerical setup or the initial formulation, and repeat the process. This reliance on continuous, expertise-intensive manual intervention limits the consistency and accessibility of topology optimization \cite{song2026adaptive, dunning2025automatic}.

Meanwhile, large language models (LLMs) have been increasingly investigated in engineering for their potential in automating physics-based analysis and computational tasks. Their relevance to these applications is supported by their general-purpose task-solving capabilities, which have been demonstrated in question answering, reasoning, code synthesis, environmental interaction, and tool manipulation \cite{zhao2023survey}. 
This human-like reasoning process of LLMs can facilitate the automation of engineering tasks by reducing the need for continuous intervention from engineers during iterative task execution and evaluation. For instance, the concept of “LLM-as-a-judge,” in which LLMs are used as evaluators, has gained attention for its potential to support evaluation processes that are typically performed by human experts \cite{gu2024survey}.
In addition to their standalone use, LLMs have been employed as autonomous agents capable of performing tasks with minimal human intervention, further broadening their applicability \cite{zhao2024expel}. Within these frameworks, LLM-based agents support problem-solving processes through capabilities such as memory, planning, and action, enabling more human-like decision-making \cite{wang2024survey}. Previous studies have explored the potential of LLM-based agents, particularly for scientific problem solving. MeLM considered forward and inverse problems in mechanics using multimodal inputs, indicating that language-model-based approaches can be applied to textual tasks, analysis and design problems involving physical data \cite{buehler2023melm}. MechGPT examined the use of LLMs in mechanics for tasks such as knowledge retrieval, language-based reasoning, and hypothesis generation, indicating that LLMs can be applied to scientific understanding and problem formulation in domain-specific contexts \cite{buehler2024mechgpt}. In computational simulation, OpenFOAMGPT demonstrated that LLM-based systems can assist with practical CFD tasks, including case setup, modification of initial and boundary conditions, changes of the turbulence model, and code translation through iterative correction \cite{pandey2025openfoamgpt}. These studies suggest that LLMs can be used not only for processing natural language but also as supporting tools for physical problem solving and computational task execution in engineering.

However, a single LLM may not always be sufficient to handle the multiple stages involved in problem solving \cite{guo2024large}, including interpreting the problem statement, using computational tools, monitoring intermediate states, and revising the solution process over multiple steps. For each stage of the decision-making process, an LLM can benefit from adopting personas with different background knowledge, each oriented toward a specific task \cite{chen2025multi}.
Multi-agent workflows have been investigated for engineering problems, with multiple LLM-based agents assuming distinct roles and interacting through subtask decomposition and verification \cite{li2024more}. MechAgents examined collaborative agent teams for elasticity and hyperelasticity problems, in which different agents were assigned roles such as planning, formulation, coding, execution, and critique, with self- and mutual-correction incorporated into the solution process \cite{ni2024mechagents}. A similar approach has also been explored in computational simulation. MCP-SIM introduced a memory-coordinated multi-agent framework that combines prompt clarification, code generation, simulation execution, error diagnosis, and explanation in an iterative loop for language-based physics simulation \cite{park2026self}. Furthermore, ALL-FEM incorporated the fine-tuned LLM in the multi-agent system, presenting a reliable and automatic simulation framework \cite{deotale2026all}.

To date, reported applications of LLM-based multi-agent systems in structural design have generally been concerned with workflows centered on concept generation, parametric modeling, and simulation-assisted iteration. DesignAgent developed a workflow for early-stage product design and evaluation, focusing on iterative concept generation and feasibility assessment \cite{chen2026llm}. DesAgent presented a multi-agent framework for mechanical design based on collaborative large and small models in conjunction with parameterized representations and engineering constraints \cite{zhang2026desagent}. Prior studies have also explored the use of the Function–Behavior–Structure ontology for the generation of controllable concepts \cite{chen2024toward}, as well as the coupling of LLMs with finite element analysis for iterative evaluation and improvement of mechanical designs \cite{jadhav2026large}. Existing work suggests that LLM-based agents can support design generation, evaluation, and refinement in engineering contexts.

Despite the aforementioned advantages of LLM-based multi-agent systems for engineering problems, their application to topology optimization specifically remains limited. A central difficulty is that topology optimization is not a single-pass generation task. After each optimization run converges, the resulting design must be assessed against the original user intent and against design-quality conventions specific to topology optimization, such as discrete material distribution and the absence of checkerboarding, and any mismatch must be addressed by revising the problem specification, the numerical parameters, or the implementation before another run is launched. Moreover, the outputs subject to evaluation include both textual and visual information, such as optimized designs and convergence plots, requiring comprehensive multimodal analysis. Assessing the multimodal data requires manual and expertise-intensive processes \cite{liu2026towards}.
Existing multi-agent workflows for engineering problems, even those that incorporate critique-style agents, primarily address errors that arise during the generation process, such as syntactic mistakes or runtime exceptions \cite{huang2023agentcoder}. However, they provide limited support for the rubric-based assessment of converged optimization results and for using such assessments to guide subsequent revisions.

To address this gap, this study introduces TopOptAgents, a multi-agent framework designed for autonomous topology optimization. The framework includes a dedicated assessment-and-revision stage in which a Critic agent inspects the converged design against the original user intent and against design-quality conventions specific to topology optimization, and issues a refinement instruction to the responsible upstream agent when the assessment is not satisfied. Together with two additional refinement loops, one operating at the specification stage and another at the code-execution stage, this post-convergence loop allows the framework to recover failures that span the full optimization workflow rather than only generation-time errors. The framework is evaluated on three benchmark topology optimization problems selected to vary how well each problem is represented in the pre-trained LLM's training distribution. This selection allows the contribution of the multi-agent self-refinement to be measured against a single-pass LLM baseline as the problem departs from cases the baseline can handle.

The remainder of this paper is organized as follows. Section \ref{sec:Multi-agent system} presents the topology optimization formulations considered in this study and introduces the proposed multi-agent framework. Section \ref{sec:Agent config} describes the roles and configurations of individual agents in detail. Section \ref{sec:results} evaluates the performance of the framework through numerical examples and examines its self-refinement capability.

\section{Multi-agent system for topology optimization problem} \label{sec:Multi-agent system}

In this section, we briefly introduce topology optimization problem considered in the present study and the proposed multi-agent system for topology optimization.

\subsection{Topology optimization}
\label{sec:topopt_form}
\subsubsection{Topology optimization problem and relevant formulations}
\label{sec:2.1.1}

\begin{figure*}[hbt!]
	\centering
	\includegraphics[scale=0.47]{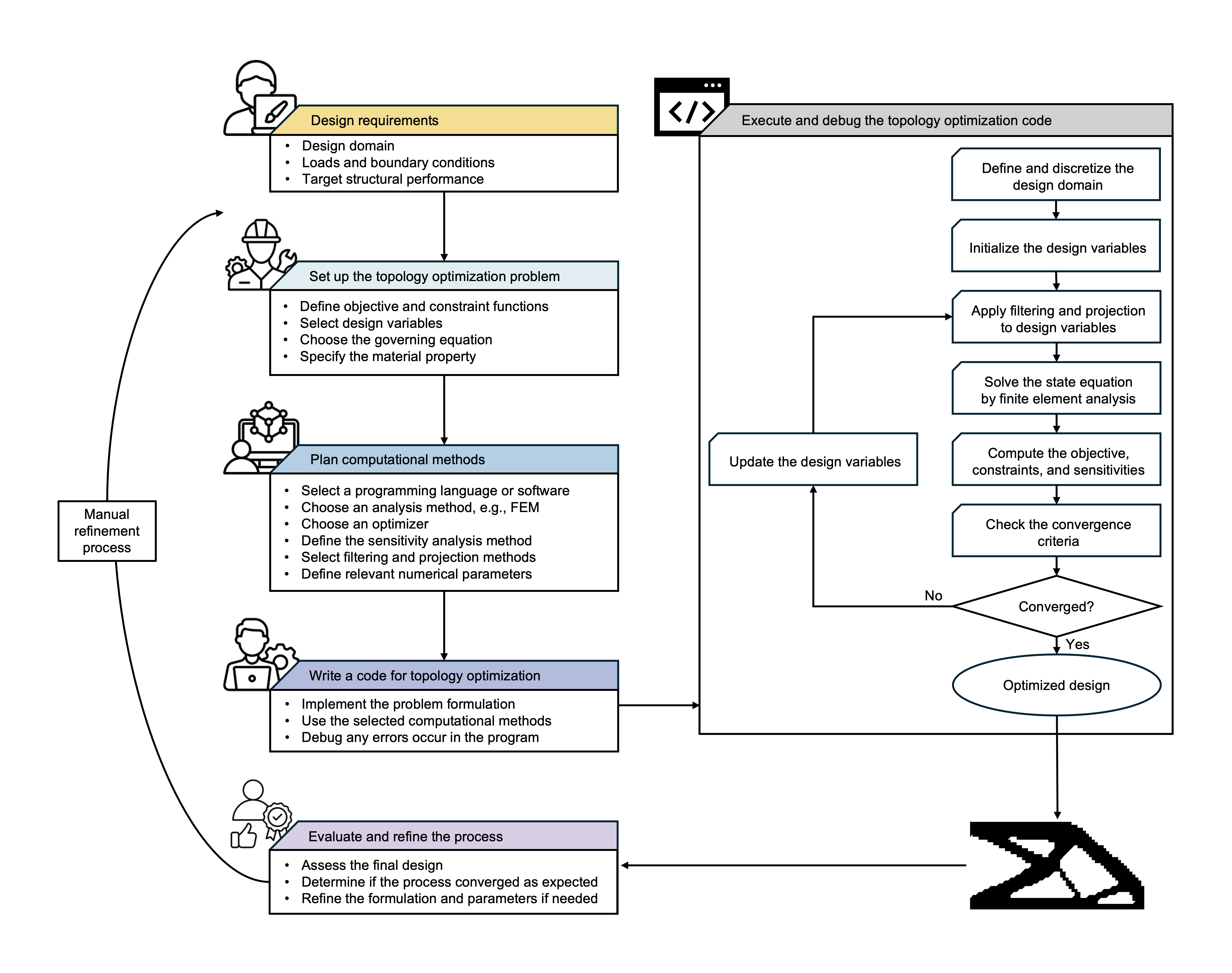}
	\caption{Overall workflow of the decision-making and computational processes for topology optimization.}
	\label{fig.1}
\end{figure*}

Topology optimization is an iterative process that involves a number of numerical methods requiring a series of decisions made by experienced engineers, as shown in Fig. \ref{fig.1}. 
The engineer first formulates a partial differential equation constrained optimization problem based on the design domain and requirements as follows:

\begin{equation}\label{eq:generalized_topopt}
    \text{subject to} \quad
\begin{aligned}
    \min_{\bm{x},\bm{s}} \quad & f(\bm{s},\bm{x}) \\
    & g_i(\bm{s},\bm{x}) \le 0, \quad i=1,\dots,m, \\
    & h_j(\bm{s},\bm{x}) = 0, \quad j=1,\dots,p, \\
    & \bm{x} \in \mathcal{X}_{\mathrm{ad}}.
\end{aligned}
\end{equation}
where $f$ denotes the objective function, $g_i$ and $h_j$ are the inequality and equality constraints, respectively, and $\mathcal{X}_{\mathrm{ad}}$ is the admissible set of design variables $\bm{x}$ \cite{sigmund2000topology}. The state variable $\bm{s}$ depends on the physics involved during the analysis, and the corresponding governing equation enters the formulation as an equality constraint. For a typical density-based compliance minimization topology optimization problem \cite{sigmund200199} that involves finite element analysis, the design variables are the element-wise material densities, ($\bm{x}=\bm{\rho}=[\rho_1, \rho_2, ..., \rho_{n_e}]^T$), and after the spatial discritization the static equilibrium takes the form $\bm{R}=\bm{K}(\bm{x})\bm{u}-\bm{F}=0$, where $\bm{K}$ and $\bm{F}$ are the stiffness matrix and the load vector. 

Since topology optimization formulation is typically gradient-based, the design sensitivities of the objective and constraint functions are required to update the design variables. Although deriving analytic sensitivity necessitates a thorough understanding of the analysis model, they are generally preferred over automated numerical differentiations to avoid computational overhead followed by automated numerical alternatives (e.g., numerical differentiation, automatic differentiation) to avoid computational overhead.
The adjoint method is typically used to evaluate the sensitivity of the objective or constraint function $\Phi$ as generalized as follows: 
\begin{equation}
    \frac{d\Phi}{d\bm{x}}
    =
    \frac{\partial \Phi}{\partial \bm{x}}
    +
    \bm{\psi}^\mathsf{T}
    \frac{\partial \bm{R}}{\partial \bm{x}}.
    \label{eq:adjoint_gradient}
\end{equation}
where the adjoint variable $\bm{\psi}$ is obtained by solving a linear adjoint equation:
\begin{equation}
    \left( \frac{\partial \bm{R}}{\partial \bm{s}} \right)^\mathsf{T}\bm{\psi}
    =
    - \left( \frac{\partial \Phi}{\partial \bm{s}} \right)^\mathsf{T},
    \label{eq:adjoint_equation}
\end{equation}

Once the mathematical formulation is determined, the engineer selects computational methods to solve the optimization problem, including the optimizer, additional techniques such as filtering and projection, and relevant numerical parameters. After all specifications are fixed, the engineer will either write code or use software to perform the iterative design search. Numerical errors often occur during execution and require refinements to the implementation.
The converged design outcome is then assessed by the engineer using metrics that vary in complexity and explicitness. For instance, convergence of the process can be verified through the flag returned by the optimization solver and the iteration log, whereas the feasibility of the material distribution (i.e., absence of checkerboard pattern) typically requires visual inspection of the design layout. Whenever the outcome fails to pass the assessment, the engineer refines the problem formulation or adjusts the parameters, and the entire process is repeated.

\subsubsection{Convergence and Quality of Optimized Results}
\label{sec:2.1.2}
\begin{figure*} [hbt!]
	\centering
	\includegraphics[scale=0.8]{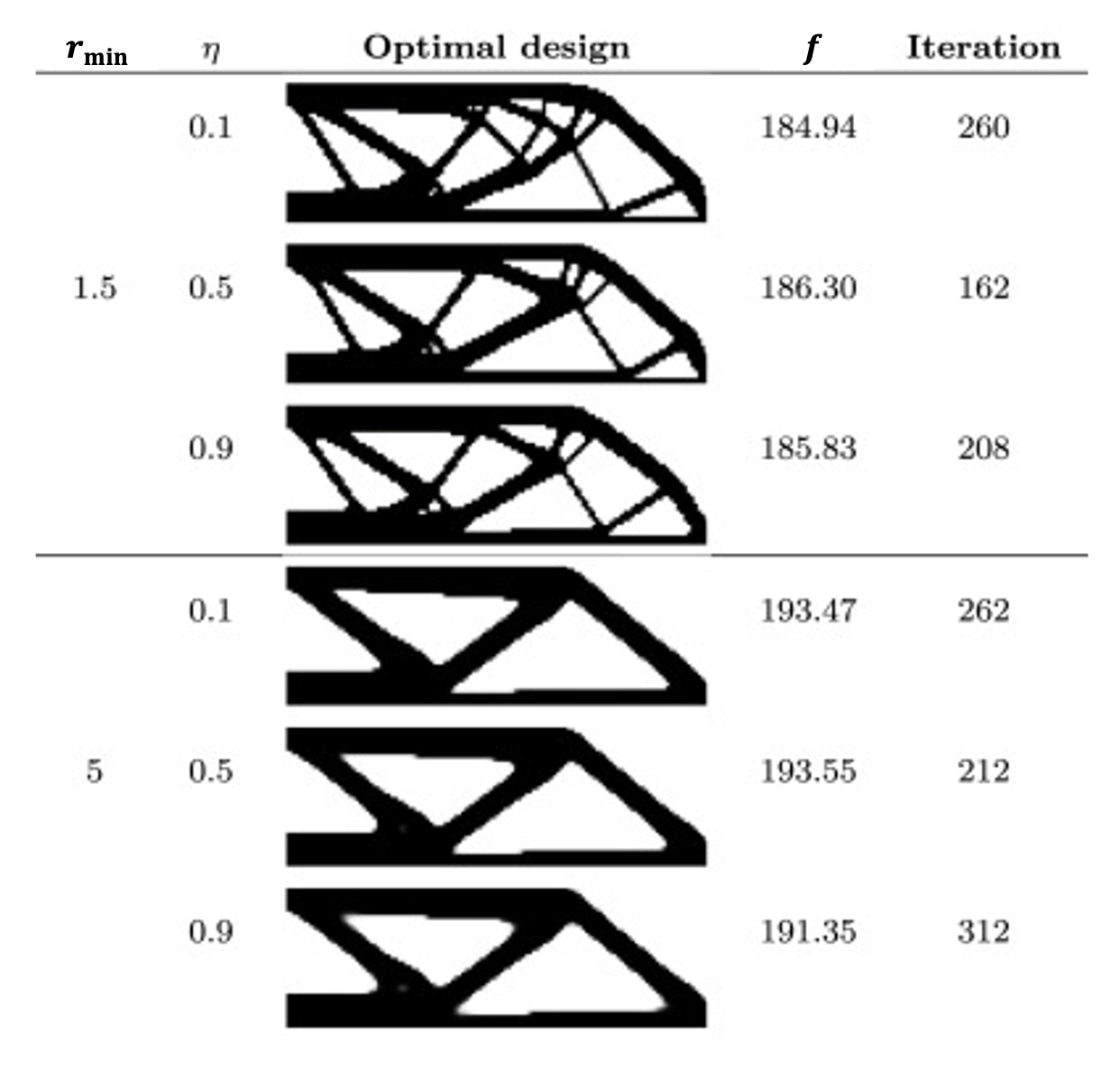}
	\caption{Effect of the filter radius {$r_{\min}$} and the projection threshold {$\eta$} on the optimized structure after Song et al. \cite{song2026adaptive}.}
	\label{fig.2}
\end{figure*}

Starting from an initial design, the optimization iteratively updates the design variables until convergence is achieved. A satisfactory optimization result is generally characterized by stable convergence of objective and constraints \cite{rojas2017short}, a clear load path with appropriate member widths, independence from mesh refinement, and smooth structural boundaries \cite{liu2014efficient}, to name a few. To achieve these qualities in practice, several numerical techniques are commonly introduced.
Optimizer-related parameters are adjusted to ensure convergence \cite{rojas2015benchmarking} and sensitivities are normalized for numerical stability. To address checkerboard patterns, mesh dependency, and intermediate density values \cite{sigmund1998numerical}, filtering \cite{lazarov2011filters} and Heaviside projection \cite{wang2011projection} are applied, with the projection sharpness $\beta$ typically increased through a continuation scheme during the optimization \cite{guest2011eliminating}.
While these techniques improve numerical robustness, they introduce additional user-defined parameters, including the filter radius $r_{\min}$, the projection sharpness $\beta$ and its continuation schedule, which are typically selected heuristically for each problem \cite{song2026adaptive}. As shown in Fig.~\ref{fig.2}, even for the same design problem, different combinations of $r_{\min}$ and $\eta$ can lead to different iteration counts, convergence behaviors, and final topologies \cite{song2026adaptive}. Taken together, the topology optimization workflow involves multiple interrelated procedures and user-defined parameters, motivating the automated decision-making framework introduced in the next section.

\subsection{Self-refining multi-agent system}

\begin{figure*} [hbt!]
	\centering
	\includegraphics[scale=0.9]{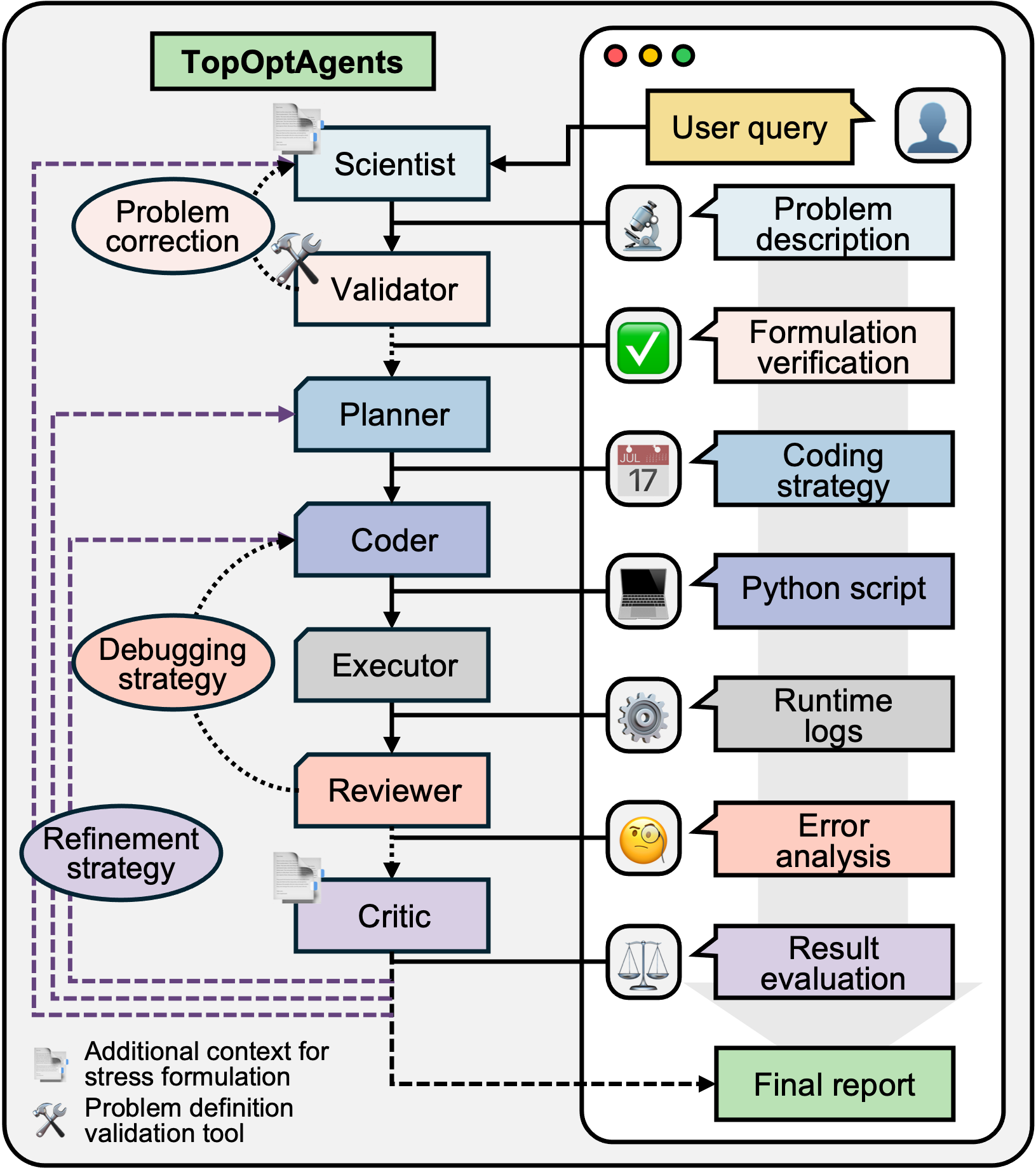}
	\caption{Overview of the TopOptAgents workflow and the outputs of each agent.}
	\label{fig.3}
\end{figure*}

In this work, we present TopOptAgents, a self-refining LLM-based multi-agent framework specially tailored to solve topology optimization problems by making a sequence of decisions required as described in Sec.~\ref{sec:topopt_form}. 
A schematic configuration of TopOptAgent is illustrated in Fig.~\ref{fig.3}, where different agents (e.g., Scientist and Validator) interact not only sequentially but also by forming subcycles (e.g., the Problem correction loop) to refine decisions made by agents. Through such, the multi-agent framework is capable of emulating and automating the decision-making process.

The overall flow of the proposed framework can be summarized as follows. Firstly, Scientist agent translates the user query describing the design problem in natural language to a topology optimization problem in the form of Eq.~\eqref{eq:generalized_topopt}. This structured problem formulation is refined through Scientist--Validator refinement loop, and Planner agent establishes a detailed plan for code generation based on the refined formulation. The plan is then passed to the code generation team, which writes a complete Python script for topology optimization. Being made up of three individual agents (Coder, Executer, Reviewer), the team iteratively revises the code until the program execution is assured and results in the optimized layout. Like a typical design process requiring manual inspection of the results, Critic agent assesses the results and writes a verdict on whether further refinement is needed. The whole process is repeated until Critic agent approves the result, making the proposed method capable of continuously improving the outcomes that were previously done by a set of experts. Through this framework, the end-to-end process from user-defined design requirements to a topology-optimized structure is decomposed into observation, planning, action, and revision stages, reducing the complexity of the tasks assigned to individual agents. This multi-agent structure reflects the practical workflow of topology optimization research. The detailed configurations of the agents are presented in Sec. \ref{sec:Agent config}.

\section{Agent configuration} \label{sec:Agent config}

All agents in the proposed workflow are configured by providing a structured system prompt to a backbone LLM model at a temperature of $0$. The system prompt for an agent is composed of five sections following \cite{oshin2025learning}: Role, Inputs, Instructions, Rules, and Output format. 
Based on the Role specified in the prompt, the agent filters its knowledge base to adopt a specific persona and define the agent’s behavior.
The Inputs section enumerates the variables passed to the agent at runtime, which may include the user request, outputs from upstream agents, and feedback from prior iterations. The input information is retrieved from the shared memory state among the agents. It is worth noting that conditional inputs can be injected to accommodate problem-specific requirements without inflating the base context.
The Instructions section prescribes the high-level reasoning procedure the agent follows to transform its inputs into the desired output, including how to prioritize among potentially conflicting sources of information. 
The Rules section enforces task-specific constraints that the output must satisfy, encoding domain conventions and expert heuristics directly into the prompt. Additional empirical rules can be introduced to prevent invalid or unexpected outputs, which are especially beneficial in directing Critic agent with a prescribed evaluation rubric.
Finally, the Output format section fixes a strict schema for the response, ensuring that the agent's output is machine-parsable and can be passed deterministically to subsequent agents in the workflow. Specifically, the agent output is strictly structured using Pydantic, guiding the agent to convey information efficiently while reducing omissions and unnecessary text.
Detailed information on the base model, input and output information, and token usage of each agent is presented in Fig. \ref{fig.5}. 
Here, the base LLM models are selected differently across agents based on the trade-off between the reasoning complexity of the task and token efficiency, as well as the need for image processing.

\begin{figure*}
	\centering
	\includegraphics[scale=0.8]{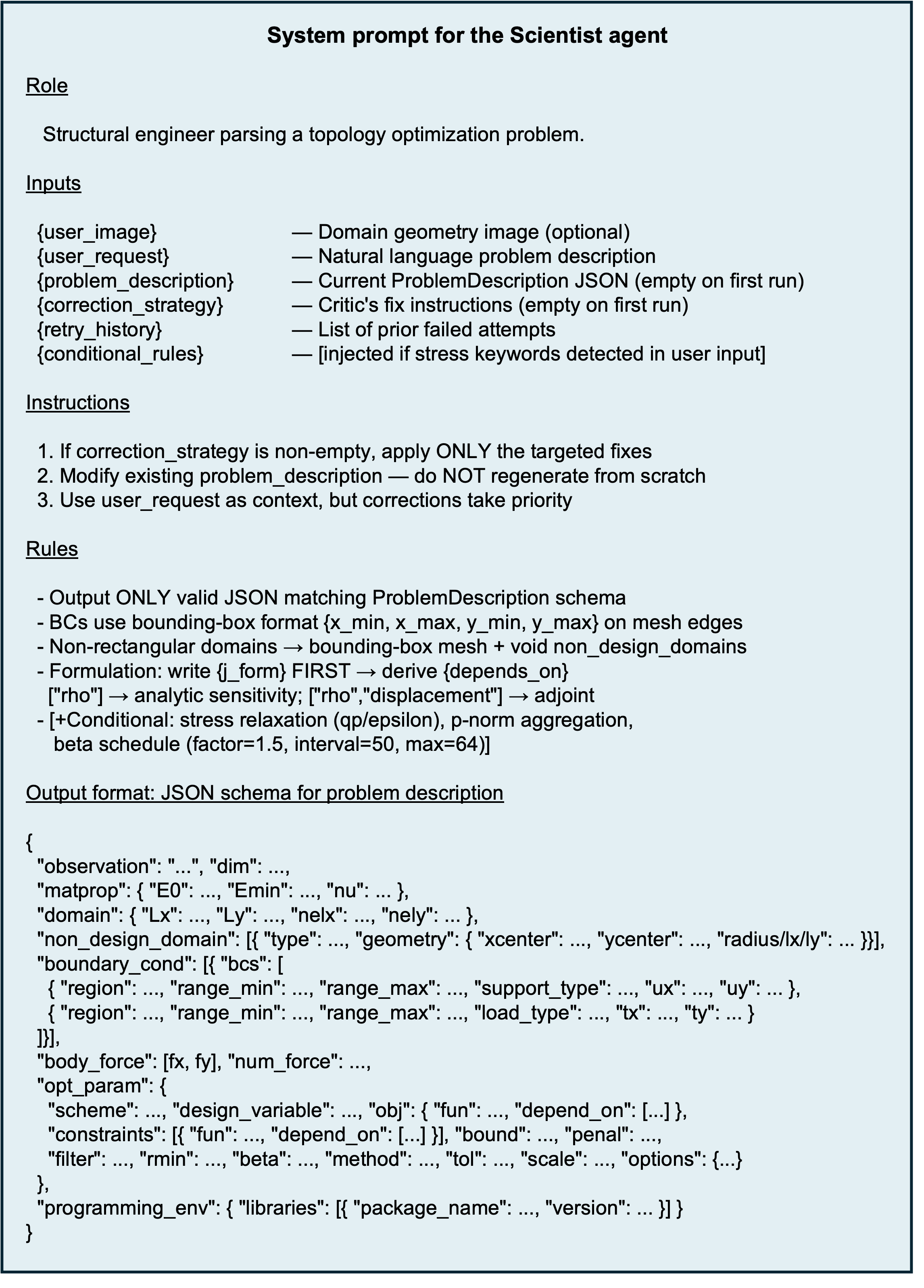}
	\caption{Example of the system prompt structure for an LLM agent represented by Scientist agent. The prompt includes role specification, input fields, instructions, rules, and output format.}
	\label{fig.4}
\end{figure*}

\subsection{Scientist agent}

Scientist agent is responsible for translating natural-language design requirements into a structured problem specification. This specification includes the mathematical formulation of the topology optimization problem in Eq.~\ref{eq:generalized_topopt}, along with the design variables, constraints, boundary conditions, and variable dependencies required for subsequent sensitivity analysis, as well as any additional numerical techniques required for implementation. Its scope is confined to problem formulation, which is required by other downstream agents. This separation of roles keeps the prompt of Scientist agent concise and prevents unnecessary inflation of the context, which may induce inconsistency in single-agent approaches as noted in \cite{park2026self}.

Within this scope, three tasks are performed by Scientist agent. 
First, it parses the user query to extract the objective and constraints functions, governing equations, and design configurations (e.g., design domain, boundary conditions). Second, it identifies both design and state variables on which the objective and constraint functions depend, which is later used by the downstream agents to determine the sensitivity analysis method. Lastly, if the user-provided query fails to specify all numerical parameters required to formulate a well-posed topology optimization problem, Scientist agent identifies the missing values and supplements them based on standard practice.

An additional key feature of Scientist's design is that its system prompt admits conditional context injection. 
Although the underlying LLM is expected to acquire considerable knowledge regarding topology optimization thanks to the widely available public open-source codes and reference \cite{sigmund200199, andreassen2011efficient}, the pre-trained model is biased to the simple problems, such as compliance minimization, and hence its context might be limited as problem complexity increases. 
To alleviate such problems, additional domain-specific context is appended to the prompt at runtime when the user query contains keywords associated with a particular kind of problem. 
This avoids inflating the base prompt with material that is irrelevant to the majority of problems while still allowing the agent to handle formulations that lie outside the LLM's pretrained knowledge. 
We demonstrate this mechanism using stress-based topology optimization as a representative case, since relevant literature and open-source code for stress formulations are comparatively limited \cite{wang2023open}. The injected context, drawn from prior studies \cite{holmberg2013stress, de2015stress, le2010stress}, summarizes stress-specific methodologies such as relaxation, p-norm aggregation, and the corresponding sensitivity treatment. The same mechanism can be extended to other problem classes (e.g., compliant mechanism design, multi-physics formulations) by preparing the corresponding context modules, without modifying the base agent.

\subsection{Validator agent}
Validator agent is designed to emulate the role of an engineer who identifies and corrects numerical problem specifications before the formulated problem is passed to the following agents. 
It examines whether Scientist's output is consistent with the user requirements and whether the problem specification contains any neglected information or potential numerical instabilities. 
If Validator fails to resolve all errors, it prompts Scientist to further refine the problem description. The two agents then go through an iterative refinement loop to revise the formulation until Validator approves it. By ensuring that the problem is well formulated before execution, Validator reduces the need for time-consuming error correction during subsequent execution and refinement.

Validator performs three different tasks: checking the consistency of Scientist's output with the user requirements, correcting errors in the boundary conditions, and identifying omissions in the problem specification. Specifically, in the first task, the agent verifies whether the objective and constraint functions, governing equations, and design configurations formulated by Scientist properly reflect the original user query. In the second task, it detects and corrects errors in the boundary conditions, such as inconsistent supports, improperly specified loads, or overlaps between boundary conditions and void regions. In the third task, it identifies missing values or unspecified numerical settings required to formulate a well-posed topology optimization problem and augments these values and settings with the filtered knowledge within LLM. For these tasks, Validator uses a predefined function that checks the existence of the required values, the aspect ratio of the mesh, and the comparison between the filter radius and the mesh size.

Furthermore, Validator seeks to mitigate numerical difficulties that arise from singularities. Validator suggests revisions based on domain-specific heuristics, such as replacing a nodal point load with an equivalent distributed load to avoid numerical singularities \cite{park2025topology}. Since such numerical singularities are typically treated based on engineers' heuristics, Validator is designed to emulate this role by applying corrective rules when possible, or by requesting further refinement from Scientist agent when the errors cannot be resolved automatically.
\subsection{Planner agent}

Once the problem definition is generated and verified through Scientist--Validator refinement loop, Planner agent writes a detailed plan for implementing the problem in a self-contained executable Python code. Planner decomposes the entire code-writing task, which is too complicated for an LLM to perform reliably in a single step \cite{pan2025codecor}, into a series of tasks for writing a single function so that each task becomes manageable. Tasks provide a step-by-step strategy for the code generation team by specifying the functions corresponding to each step of the topology optimization workflow described in Fig.~\ref{fig.1}, the parameter initialization based on Scientist's specifications, the input and output variables of each function, the way that the numerical formulations are translated into a computational language, and the configuration of the iterative loop for topology optimization. Since the plan is based on Scientist's problem specification, Planner can reduce the prompt tokens required to include the full theoretical background while maintaining consistency with Scientist's problem formulation.

\subsection{Code Generation Team}
Three agents iteratively collaborate to produce a single Python script for topology optimization. To maintain the continuity of the workflow, the generated code is expected to run without runtime errors and produce the final material layout and convergence log of the topology optimization process. Coder first writes a self-contained program, and Executor then runs the code through a subprocess. If the expected output is not obtained or an error occurs, Reviewer sends the process back to Coder for error correction. Through this iterative correction process, the code generation team ensures that the final material layout and convergence log are successfully generated for subsequent analysis by Critic.

Coder agent writes a code for topology optimization based on Planner agent’s task list. Its system prompt includes a one-shot example in the form of a skeleton Python script for cantilever-beam compliance minimization. In this skeleton code, finite element analysis is performed using the FEniCS library \cite{alnaes2015fenics}, and the optimization process is handled by pyOptSparse \cite{wu2020pyoptsparse}. In particular, FEniCS is used as the computational basis for mathematical analysis, providing finite element tools for solving partial differential equations. It also supports the computation of partial and adjoint sensitivities, facilitating sensitivity analysis within the proposed architecture. Guided by the skeleton code, Coder agent generates a script that implements the full topology optimization workflow. Executor agent is a non-LLM agent responsible for executing the program generated by Coder agent. The resulting outputs, including error messages, final-structure visualizations, and convergence plots, are then passed to Reviewer agent. Reviewer agent analyzes the execution results and determines whether the execution was successful. When errors arise during execution, Reviewer agent invokes Coder agent with a correction strategy, prompting it to revise the program until it runs successfully. These three agents create a self-correcting loop that continues until the program runs without errors and produces all expected outputs.
\subsection{Critic agent}

Critic agent plays a key role in orchestrating the TopOptAgents and refinement loops. The agent evaluates the overall execution results using a self-created rubric for multimodal outputs, including textual (e.g., convergence history) and visual information (e.g. design layout). It is designed to comprehensively assess the optimization outcome based on the rubric and to determine whether further refinement is required. When the result does not satisfy the evaluation criteria, Critic agent identifies which issue should be addressed first and which agent should be engaged next. The corresponding refinement strategy is then passed to the selected agent to guide how the execution should be adjusted and rerun. 
This refinement loop continues until Critic agent approves the result. If all criteria are satisfied, the multi-agent workflow terminates, and the final report is provided in the user's preferred language. The user may also provide additional feedback on the result, which triggers another iteration beginning with Scientist agent.

The evaluation criteria are organized into four articles according to the priority of refinement.
The first criterion is \textit{output validity}, which checks whether the generated files are non-empty and non-trivial. The second criterion is \textit{problem formulation consistency}. Under this criterion, Critic agent reviews whether the objective function, constraint functions, boundary conditions, and sensitivity formulations are consistent with the intended problem definition, including their discretized forms. The third criterion is \textit{convergence}. Based on the visual convergence plots and the output flags from the optimizer, Critic agent assesses the adequacy of the iteration count, the progress of the optimization, and the solver termination status. In particular, with the capability of perceiving multimodal data including images, Critic can examine the design and convergence history more efficiently, which would otherwise be token-intensive due to the amount of data generated during the iterative optimization process. The final criterion is \textit{design quality}, which evaluates the structural discreteness and material connectivity of the generated design. In this step, geometric and numerical issues are inferred from the visual outcomes, supported by the textual optimization logs. Although the LLM is aware of heuristic assessment criteria, additional guidance can be provided to focus Critic agent's assessment on certain rubric items. For example, in density-based topology optimization, this guidance may include gray regions, checkerboard patterns, and overly thin members, which may indicate poor convergence or physically unreliable layouts. Through the joint interpretation of textual execution logs, numerical convergence behavior, and visual design features, the Critic agent evaluates the four criteria and either accepts the outcome or formulates a refinement strategy based on the highest-priority criterion that is not satisfied.

\begin{figure*}
	\centering
	\includegraphics[scale=0.55, angle=90]{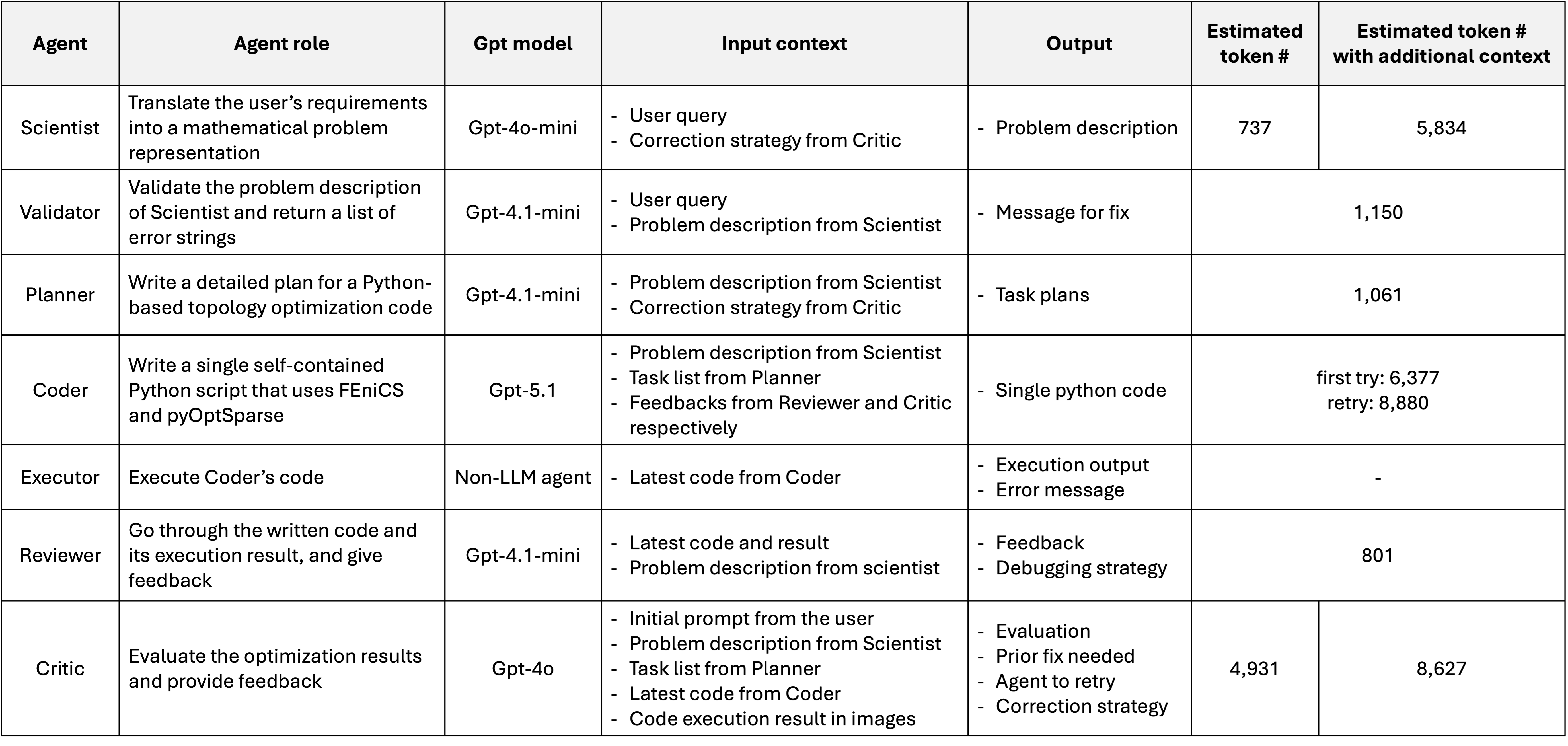}
	\caption{Profiles of the agents in TopOptAgents.}
	\label{fig.5}
\end{figure*}

\section{Results and Discussions} \label{sec:results}

This section demonstrates that the multi-agent self-refinement framework provides measurable benefit on topology optimization problems by evaluating three benchmark problems chosen to span a range of training-distribution coverage. We compare TopOptAgents against a single-pass baseline LLM and observe that the multi-agent self-refinement recovers the success rate where the baseline's performance degrades. To illustrate the workflow, we first trace a successful execution that completes without any refinement (Sec.~\ref{sec:simple_exec}), and then examine three refinement loops that operate at distinct stages of the pipeline and recover structurally distinct failure modes. Finally, we describe the components through which the user receives the workflow's outcome and initiates a further refinement cycle.

\subsection{Benchmark problems} \label{sec:benchmark}

Three benchmark topology optimization problems are considered herein: compliance minimization for a cantilever beam, compliance minimization for an MBB beam with the load applied at the midpoint of the right edge, and stress minimization for an L-shaped beam. It is worth noting that these three problems were chosen to vary the degree to which similar formulations and code appear in the pre-trained LLM's training distribution; pre-trained LLMs are known to exhibit a quantitative bias toward examples that are frequently represented in their training corpus, performing systematically better on tasks and code variants that resemble those memorized examples than on otherwise comparable but less-represented ones \cite{li2024task,riddell2024quantifying}. 

\begin{figure*}
	\centering
	\includegraphics[scale=0.65]{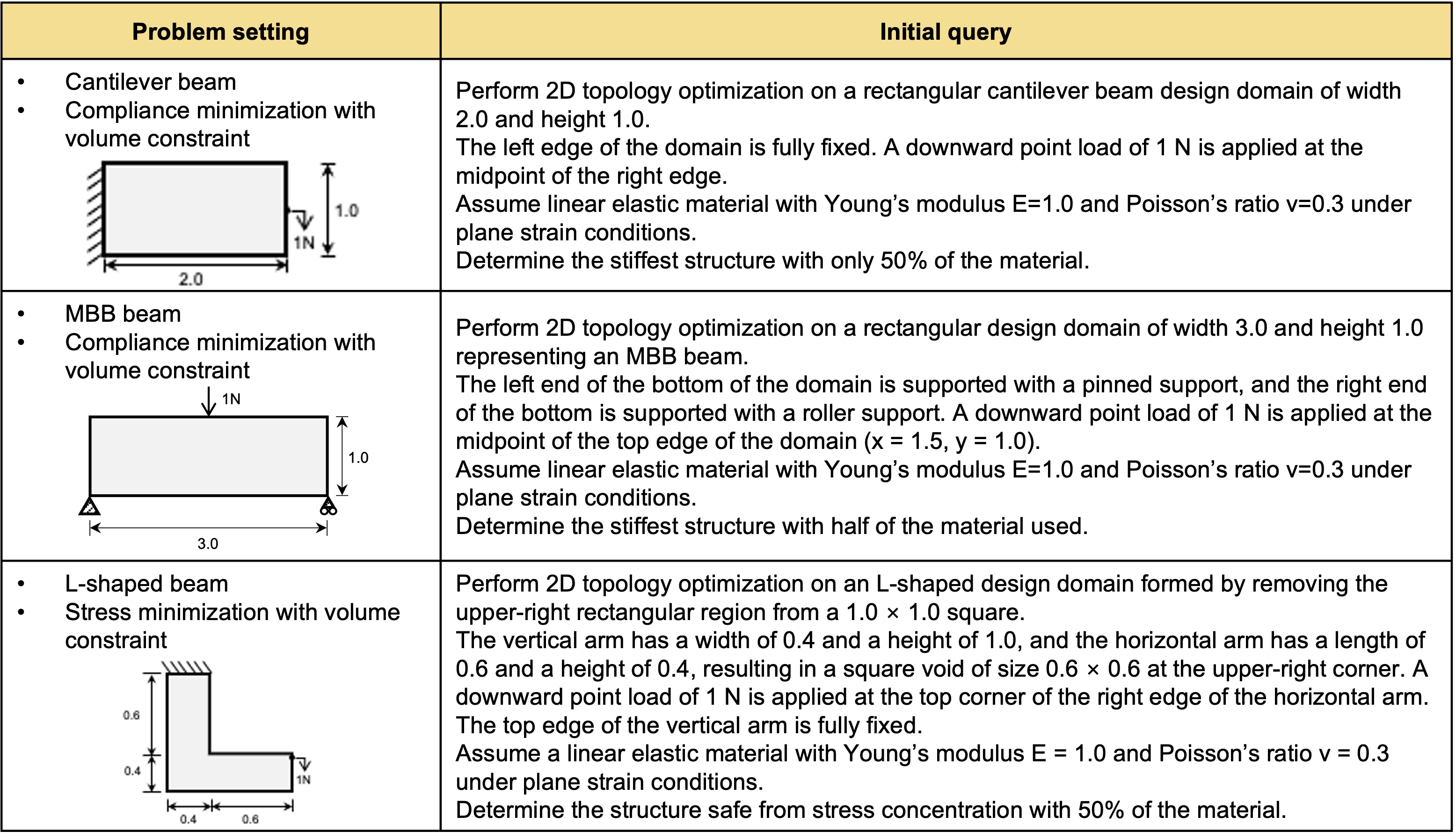}
	\caption{Benchmark problem settings and initial user queries used for evaluating TopOptAgents.}
	\label{fig.6}
\end{figure*}

The cantilever bending problem is the most widely covered case, as it is a canonical compliance-minimization example used throughout topology-optimization tutorials and is closely paralleled by widely disseminated open-source implementations such as the MATLAB top88 code \cite{sigmund2022benchmarking}. The MBB beam problem is moderately covered; the general problem formulation is well established, but the load-position variant considered here, in which the load is relocated from the lower-right corner to the midpoint of the right edge, is rarely documented in this specific form. Therefore, the system should combine recalled domain knowledge with appropriate boundary-condition adaptation rather than reproduce a memorized example. The L-shaped stress minimization problem is the least covered case. Stress-driven topology optimization with constraint aggregation, such as p-norm aggregation, is comparatively scarce in publicly available code and tutorials. The L-shaped domain further introduces a localized geometric stress singularity at the re-entrant corner, making the configuration susceptible to numerical inaccuracies even under subtle changes in the problem setup \cite{duysinx1998topology}. Thus, neither a memorized example nor a straightforward adaptation of a compliance template is sufficient. The corresponding user prompts and domain specifications are summarized in Fig. \ref{fig.6}.

\subsection{Comparison with a single baseline LLM} \label{sec:baseline_compare}

To assess where the multi-agent self-refinement workflow provides measurable benefit over a single-LLM baseline, we compared TopOptAgents against GPT-5.4 Thinking, a commercially available reasoning model used as a single-pass baseline at the time of evaluation (April 2026). Each method was evaluated over 10 independent sessions for each problem. A trial was counted as successful if the optimization converged to a valid design consistent with the problem setup. Failures included non-convergence, numerical instability, mismatch with the user's initial query, and code-level errors.

\begin{figure*}
	\centering
	\includegraphics[scale=0.8]{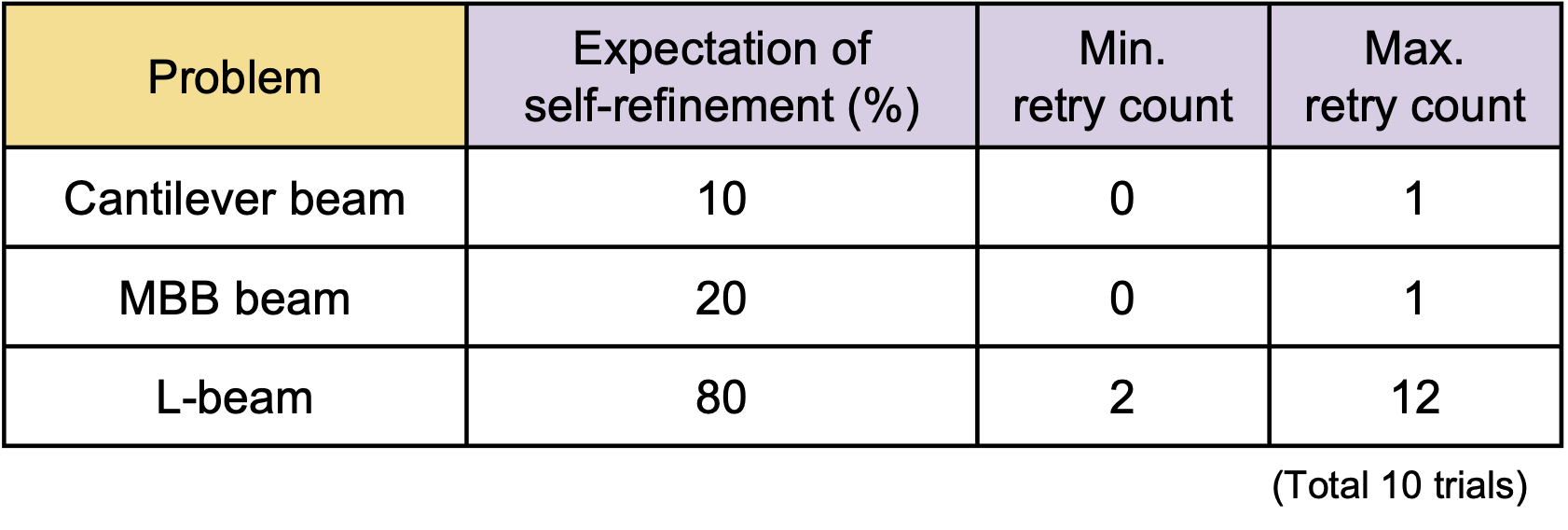}
	\caption{Summary of expected self-refinement and retry counts for the benchmark problems.}
	\label{fig.7}
\end{figure*}

Figure~\ref{fig.7} summarizes the number of self-refinement iterations used by TopOptAgents on each problem. The iteration count tracks the coverage variation identified in Sec.~\ref{sec:benchmark}: zero refinement is required on the cantilever problem, modest refinement on the MBB variant, and up to 12 refinement iterations on the L-shaped stress problem to pass the test. This pattern reflects the additional reasoning steps that the framework performs in cases where a single LLM cannot complete the problem in one pass. The number of refinement iterations should therefore be interpreted not as an isolated computational cost to be minimized, but as a direct measure of the additional reasoning required by the framework for each problem.

\begin{figure*}
	\centering
	\includegraphics[scale=0.7]{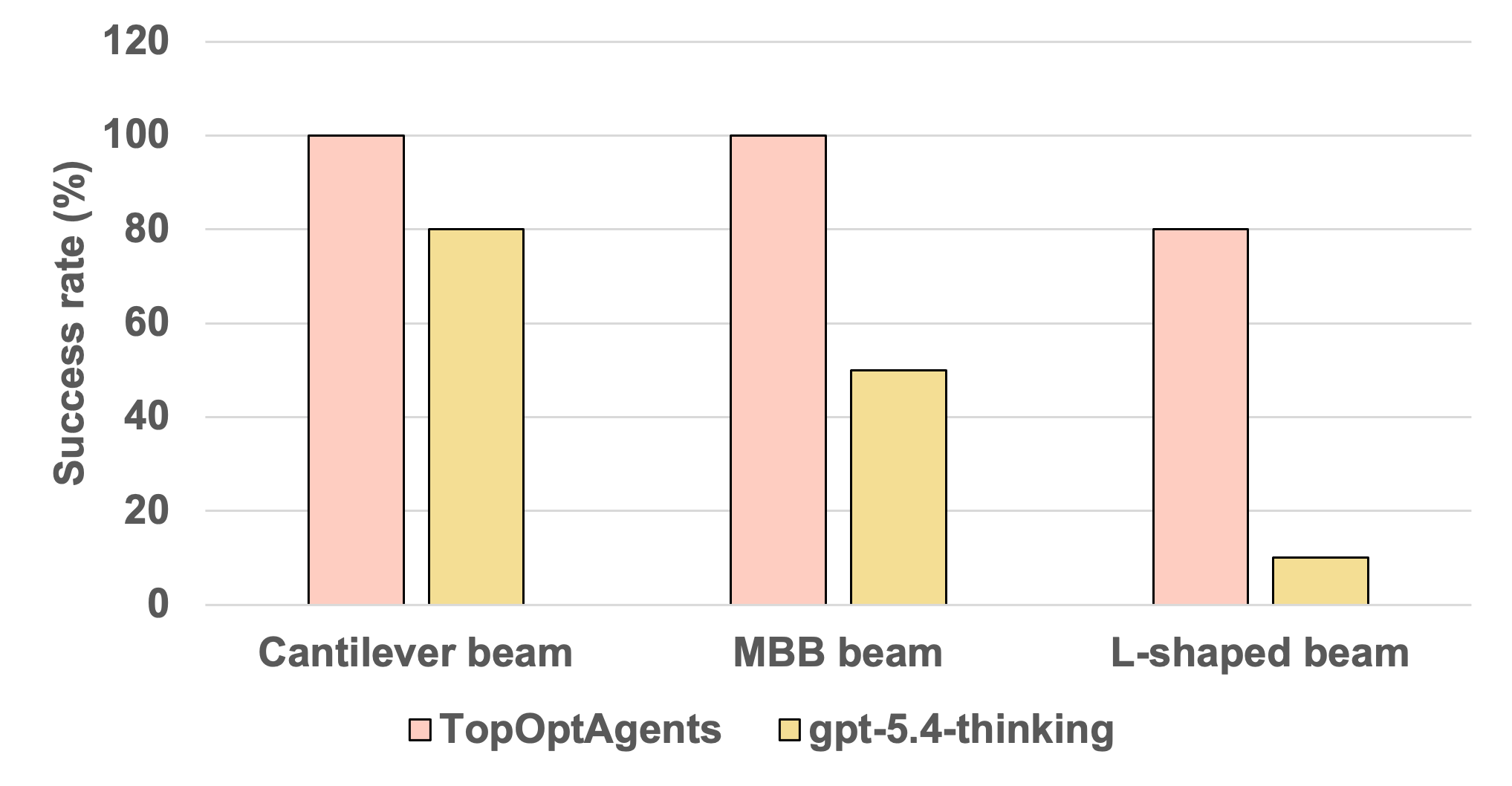}
	\caption{Success rates over 10 trials for three benchmark topology optimization problems, comparing TopOptAgents and gpt-5.4-thinking of OpenAI.}
	\label{fig.8}
\end{figure*}

The success rates of each method over the 10 trials are also compared in Figure~\ref{fig.8}; the success rate difference between the multi-agent system and the single LLM baseline becomes more pronounced as the problem diverges from the basic cantilever beam problem. The improvement from 80\% to 100\% on the MBB variant shows that even a mild departure from the memorized example can reduce baseline success by 20 percentage points. The Validator--Scientist and Reviewer--Coder loops (Sec.~\ref{sec:val_sci}, Sec.~\ref{sec:reviewer}) recover these failures by catching boundary-condition inconsistencies and code-level errors arising from the load-position variant. The improvement from 10\% to 80\% on the L-shaped problem shows where the framework provides the largest gain: the observed baseline failure modes, including inconsistent boundary conditions, singular finite element systems, and unstable stress-driven optimization, occur across multiple stages of the workflow. It can therefore be concluded that including the post-execution Critic stage (Sec.~\ref{sec:critic}) can detect converged-but-wrong outputs that are not exposed by a single-pass workflow.

\subsection{Multi-agent topology optimization without self-refinement} \label{sec:simple_exec}

\begin{figure*}
	\centering
	\includegraphics[scale=0.55]{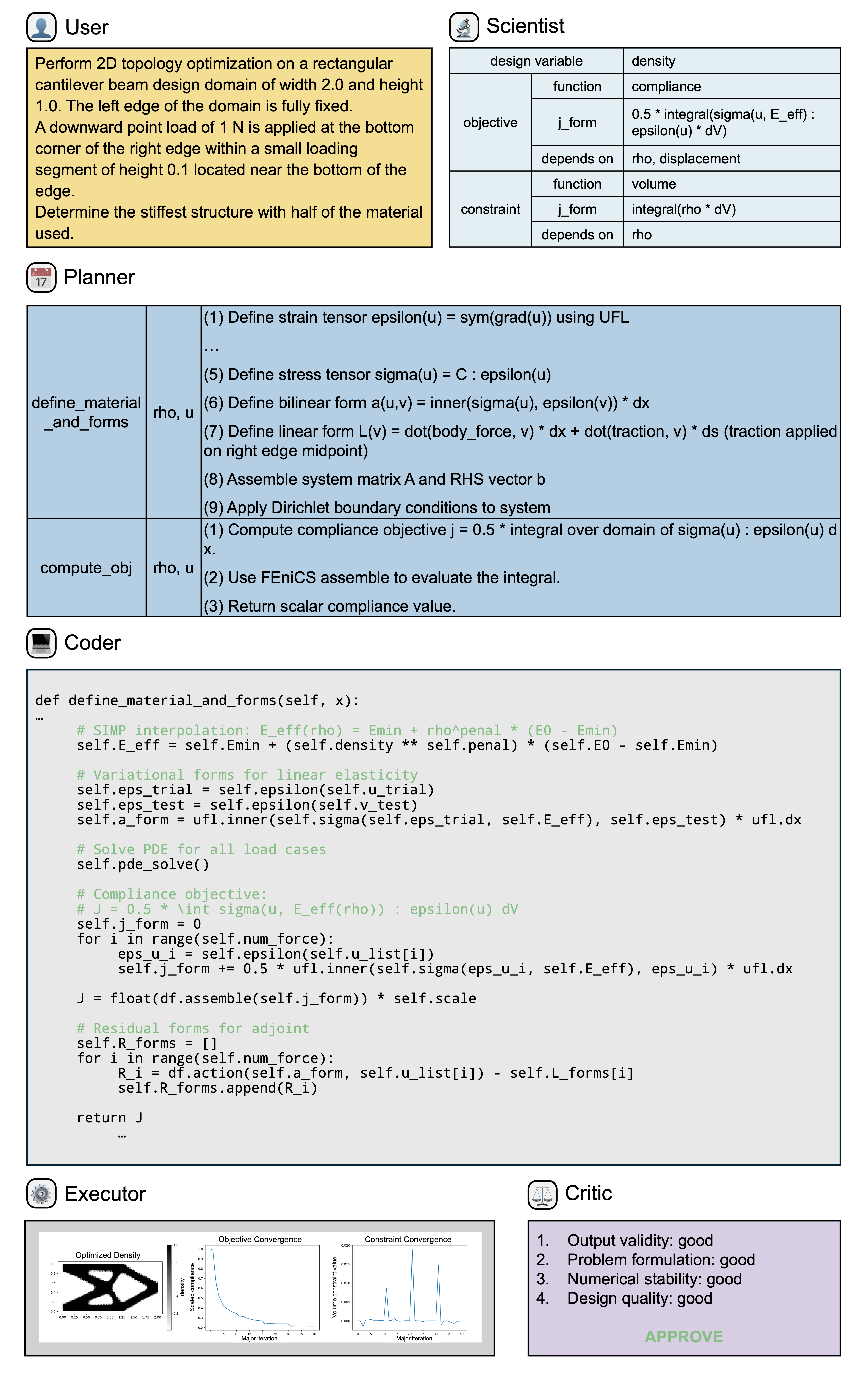}
	\caption{Overall execution flow of TopOptAgents for the cantilever beam compliance minimization problem.}
	\label{fig.9}
\end{figure*}

For the compliance minimization problem with a cantilever-beam configuration, the proposed multi-agent workflow obtains a valid optimized design without invoking any self-refinement loop. Figure~\ref{fig.9} traces the execution sequence agent by agent and provides the baseline against which the refinement cases of Sec.~\ref{sec:self-refinement} are compared.

The user query specifies the design objective in natural language as ``the stiffest structure", which Scientist translates into a formal optimization problem including compliance as the objective function, volume fraction as the inequality constraint, and the static linear elasticity governing equation. Scientist additionally identifies the dependencies of each function on the design variable and the state variable, which the downstream agents use to determine the sensitivity formulation.

Validator inspects Scientist's specification for consistency with the user query, for parameter completeness, and for known numerical stability conditions, including the mesh aspect ratio and the relationship between the filter radius and the element size. For the cantilever case, all of these checks pass on the first inspection, so the Validator--Scientist refinement loop is not activated. Planner then decomposes the implementation into a task list of single-purpose functions covering mesh and boundary condition construction, finite element assembly, sensitivity computation, density filtering, projection, and the iteration loop for the optimization process.

The code generation team executes Planner's task list: Coder produces a self-contained Python script that uses FEniCS for the finite element analysis and pyOptSparse with SNOPT \cite{gill2005snopt} as the optimizer; Executor runs the script as a subprocess; Reviewer inspects the run for runtime errors and missing output artifacts. The cantilever-case script runs without error on its first execution, so the Reviewer--Coder refinement loop is not invoked either. The optimizer converges to a discrete material distribution that minimizes compliance under the prescribed volume-fraction constraint.

Critic finally evaluates the output against the four-criterion rubric described in Sec.~\ref{sec:Agent config}: output validity, problem-formulation consistency, convergence behavior, and design quality. All four criteria are satisfied on the first evaluation, so the Critic-driven system-level refinement loop is not triggered, and the workflow terminates with a report (Sec.~\ref{sec:report}). The fact that none of the three refinement loops is needed on this problem is consistent with the success-rate observation in Sec.~\ref{sec:baseline_compare}: the cantilever compliance problem has the highest training-distribution coverage among the three benchmarks, and a single-pass LLM solution is sufficient on its own.

\subsection{Self-refinement cases} \label{sec:self-refinement}

The three refinement loops in TopOptAgents are organized to catch and autonomously correct structurally distinct failure modes in distinct stages of the pipeline when solving the MBB beam and L-shaped beam optimization problems. The Validator--Scientist loop (Sec.~\ref{sec:val_sci}) operates on the problem specification before any code is written or executed. It catches specification-level failures such as inconsistent boundary conditions or omissions in the numerical setup.
The Reviewer--Coder loop (Sec.~\ref{sec:reviewer}) operates during execution, on the runtime signals generated by the code. It catches implementation-level failures such as syntax errors, missing library functions, or unhandled execution paths that would otherwise prevent successful completion.
The Critic-driven system-level loop (Sec.~\ref{sec:critic}) operates after a converged solution is produced. It catches semantic-quality failures that pass both prior stages,
such as a converged design whose constraint formulation or boundary conditions diverge from the user's intent, or a converged design with poor discreteness or checkerboarding due to under-adjusted numerical parameters. Because each loop is the only mechanism in the workflow with access to its corresponding stage of information, a failure mode that originates at one stage cannot be reliably corrected by a loop operating at another stage.
The two benchmark problems exercise these loops to different degrees: the MBB compliance case activates loops at the specification and implementation stages to handle the load-position variation, whereas the L-shaped stress case requires the deepest sequence of corrections involving all three loop types.

\subsubsection{Validator-Scientist loop} \label{sec:val_sci}

The Validator--Scientist loop is the only refinement mechanism in the workflow with access to the problem specification before any code is written. It operates in two intervention modes: direct correction by Validator when the inconsistency is local and resolvable from the specification alone, and a requested reformulation back to Scientist when the issue requires reasoning beyond Validator's corrective rules. The following two examples illustrate each mode.

\begin{figure*}
	\centering
	\includegraphics[scale=0.85]{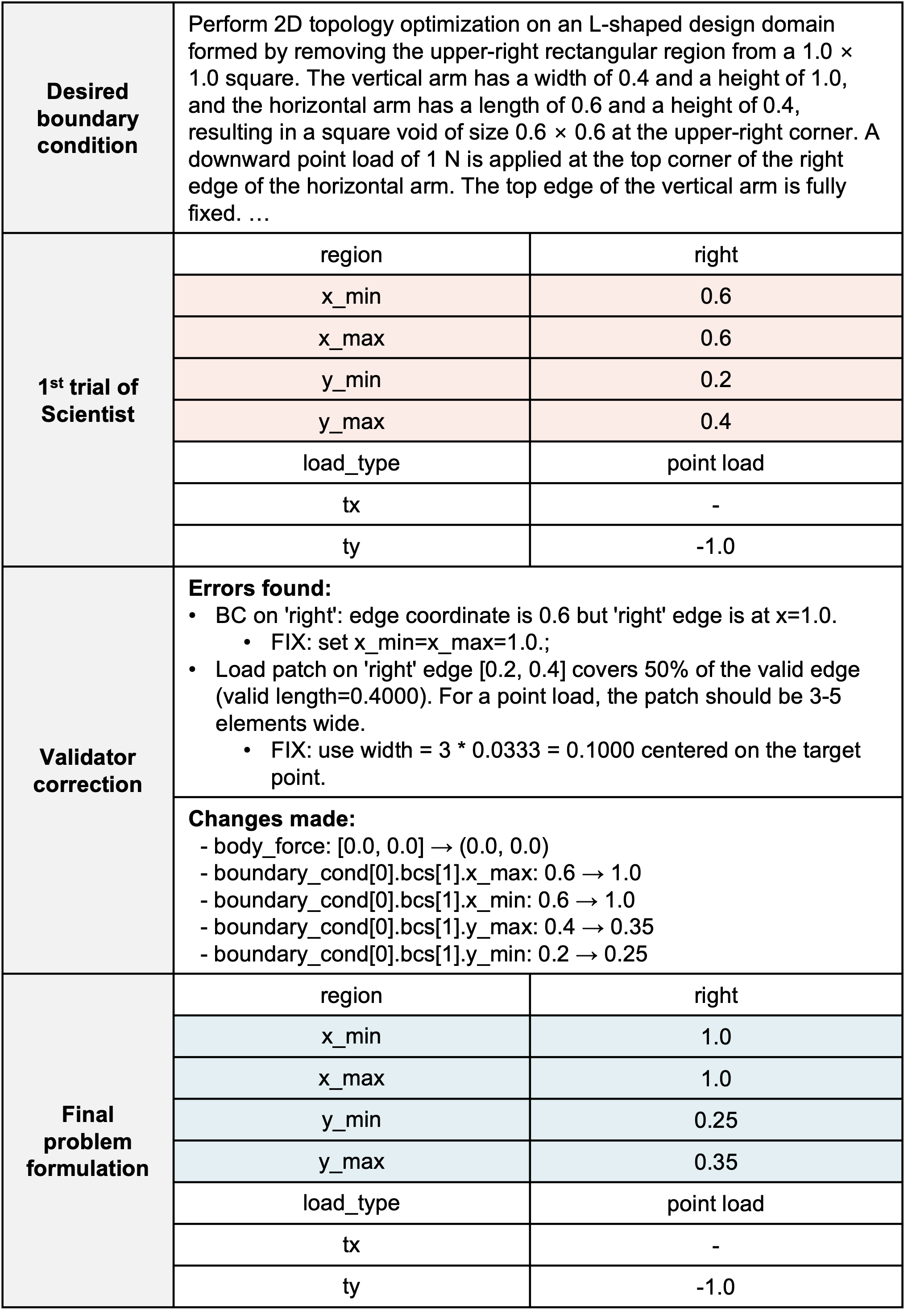}
	\caption{Example of Validator agent directly correcting an incorrectly specified external load location in Scientist’s initial problem description.}
	\label{fig.10}
\end{figure*}

Figure~\ref{fig.10} shows the direct correction mode by the loop. Scientist initially placed the $x$-coordinate of the applied load at the middle of the horizontal arm of the L-shaped domain, whereas the user's query had specified the load at the upper-right end of the arm. Validator detected this mismatch against the user query and shifted the load to $x = 1.0$, corresponding to the right end of the arm. The correction was applied directly without re-invoking Scientist, and the specification proceeded to Planner.

\begin{figure*}
	\centering
	\includegraphics[scale=0.9]{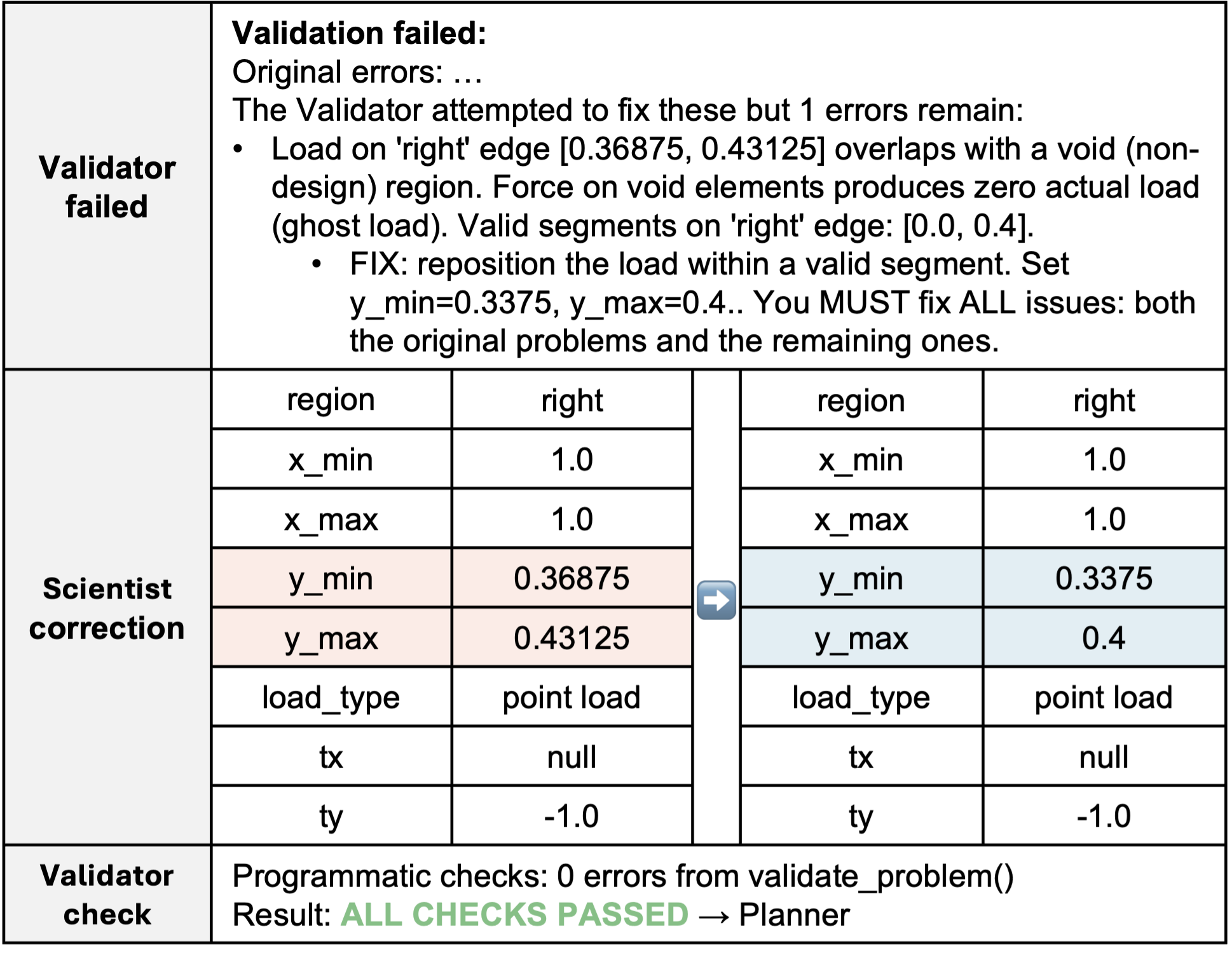}
	\caption{Example of Validator agent requesting reformulation when the specified external load location overlaps with a void region.}
	\label{fig.11}
\end{figure*}

Figure~\ref{fig.11} shows the reformulation mode, triggered by a specification produced by Scientist that placed the external load within a region designated as void, producing an ill-posed finite element system. Resolving this conflict requires deciding how to relocate either the load or the void, a refinement of the problem formulation that falls within Scientist's scope rather than Validator's rule-based corrections. The loop continues until a consistent specification is produced. Together, these two examples define the scope of the Validator--Scientist loop: it corrects simple specification-level inconsistencies in place and escalates structurally ambiguous cases back to Scientist before execution begins.

\subsubsection{Reviewer feedback within the code generation team} \label{sec:reviewer}

The Reviewer--Coder loop operates within the code generation team (Coder, Executor, Reviewer), where Reviewer examines execution outcomes and returns a correction strategy to Coder. It is the refinement mechanism in the workflow with access to runtime error signals. Failures that surface only during execution, such as undefined library functions, type mismatches, or invalid argument lists, cannot be diagnosed from the specification alone and would otherwise prevent successful execution. Its scope is limited to code correctness; design quality and problem-formulation issues are handled by Critic (Sec.~\ref{sec:critic}) and the Validator--Scientist loop (Sec.~\ref{sec:val_sci}), respectively. It is worth noting that a similar code-generation team architecture, with agents dedicated to writing, executing, and reviewing code in an iterative loop, has previously been used in language-based physics simulation \cite{park2026self}. The following two examples illustrate both the typical correction path and the robustness of the loop to imperfect Reviewer feedback.

\begin{figure*}
	\centering
	\includegraphics[scale=0.6]{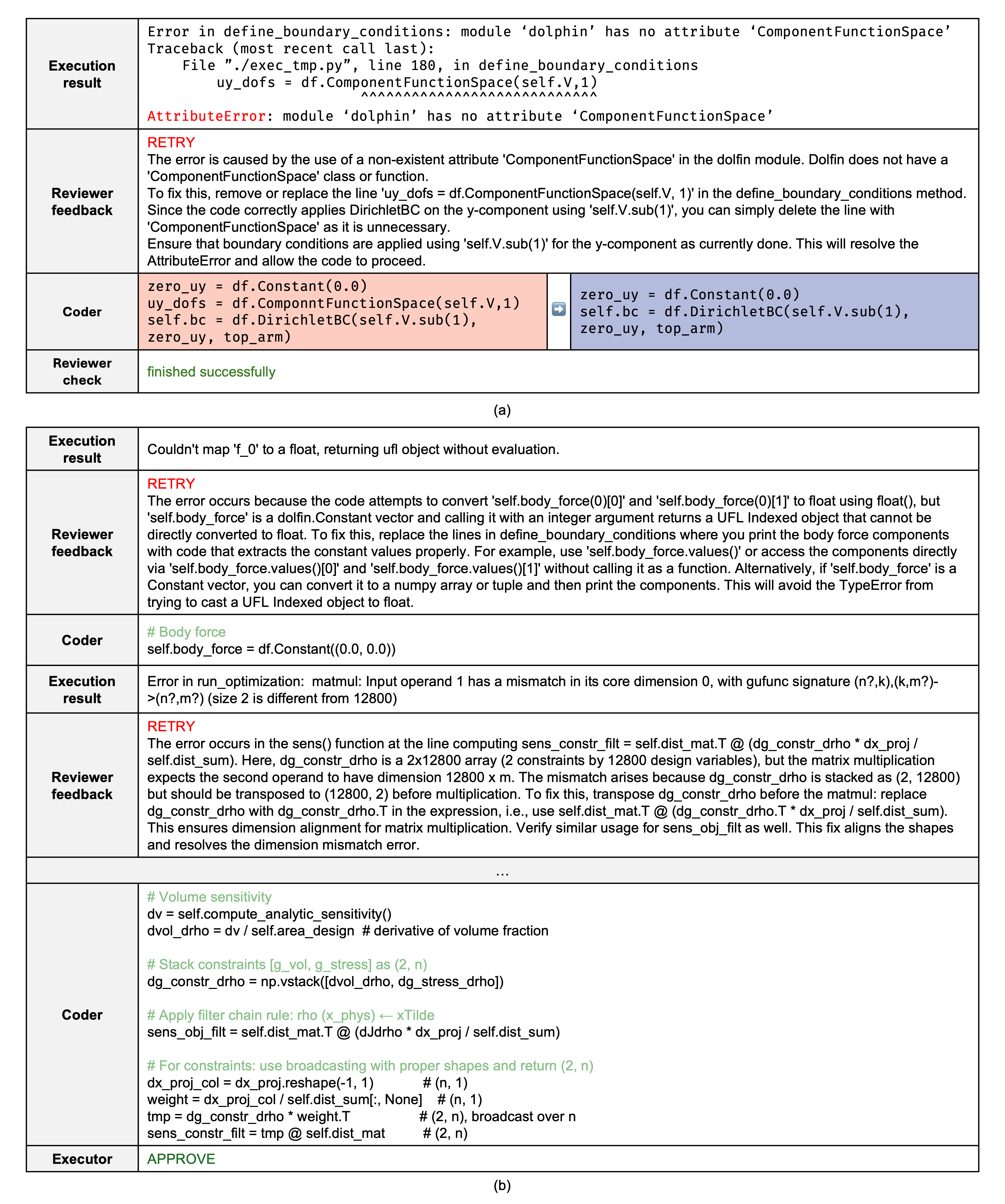}
	\caption{(a) Example of code-level error correction by Reviewer agent within the code generation team. (b) Example of iterative code refinement within the code generation team to obtain an executable program.}
	\label{fig.12}
\end{figure*}

Figure~\ref{fig.12}(a) shows the typical correction path. The initial program written by Coder called a function that is not part of the FEniCS library, which caused the run to fail. Reviewer parsed the error message, identified that the function was unnecessary for the intended computation, and returned this diagnosis to Coder. Coder removed the corresponding line, and the program executed successfully on the next run.

Figure~\ref{fig.12}(b) shows that the loop tolerates imperfect Reviewer feedback. In this case, the correction Reviewer suggested introduced a new error rather than resolving the original one. Because the same loop is re-entered on every failed execution, the additional error was caught on the next iteration and a corrected version of the program was eventually produced. The robustness exhibited here is a property of the multi-agent composition rather than of any single agent.

\subsubsection{Critic-driven refinement at the system level} \label{sec:critic}

\begin{figure*}
	\centering
	\includegraphics[scale=0.55]{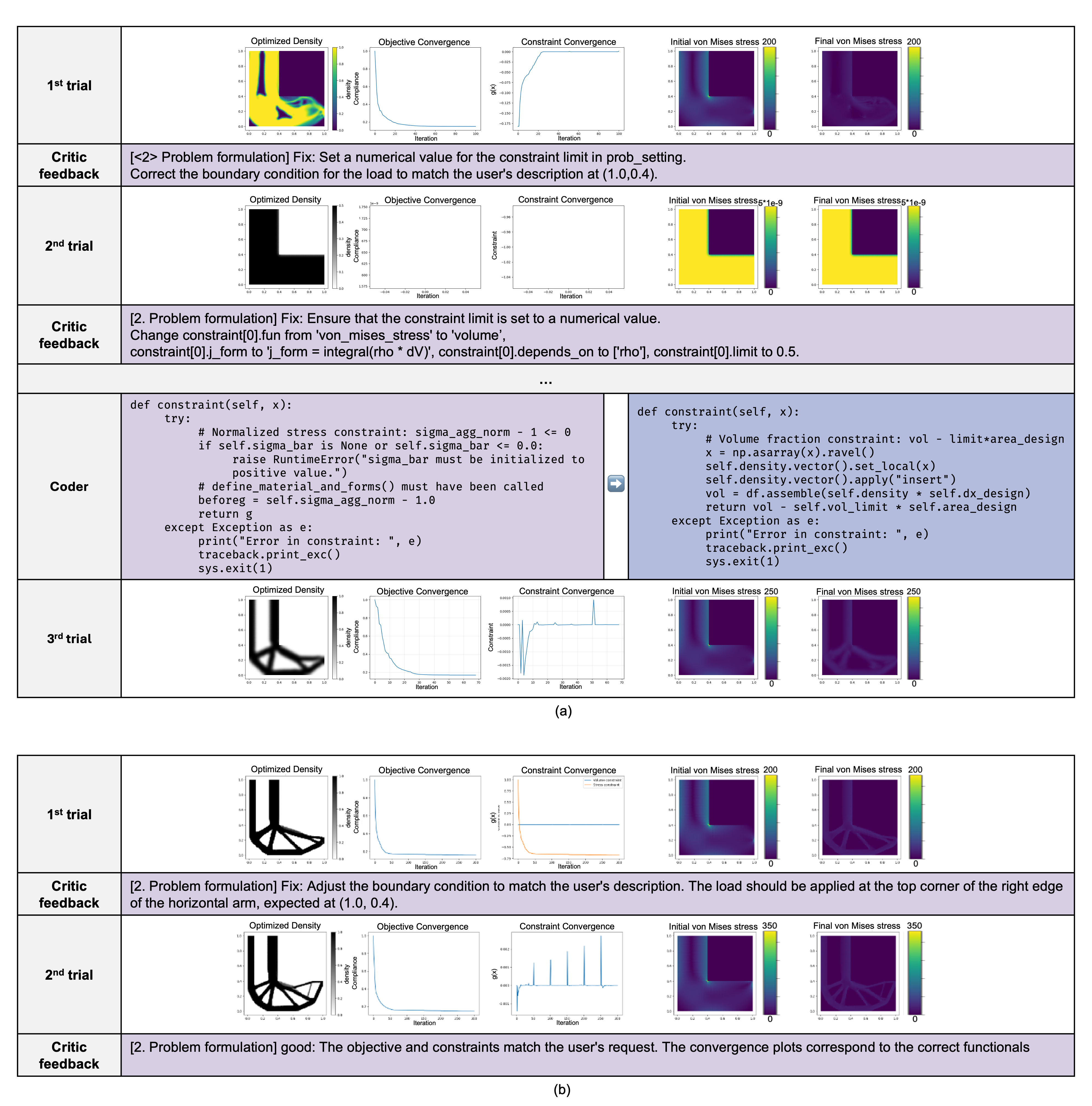}
	\caption{(a) Example of Critic agent identifying an incorrect choice on the constraint function. (b) Example of Critic agent correcting a force location inconsistent with the user’s request.}
	\label{fig.13}
\end{figure*}

\begin{figure*}
	\centering
	\includegraphics[scale=0.85]{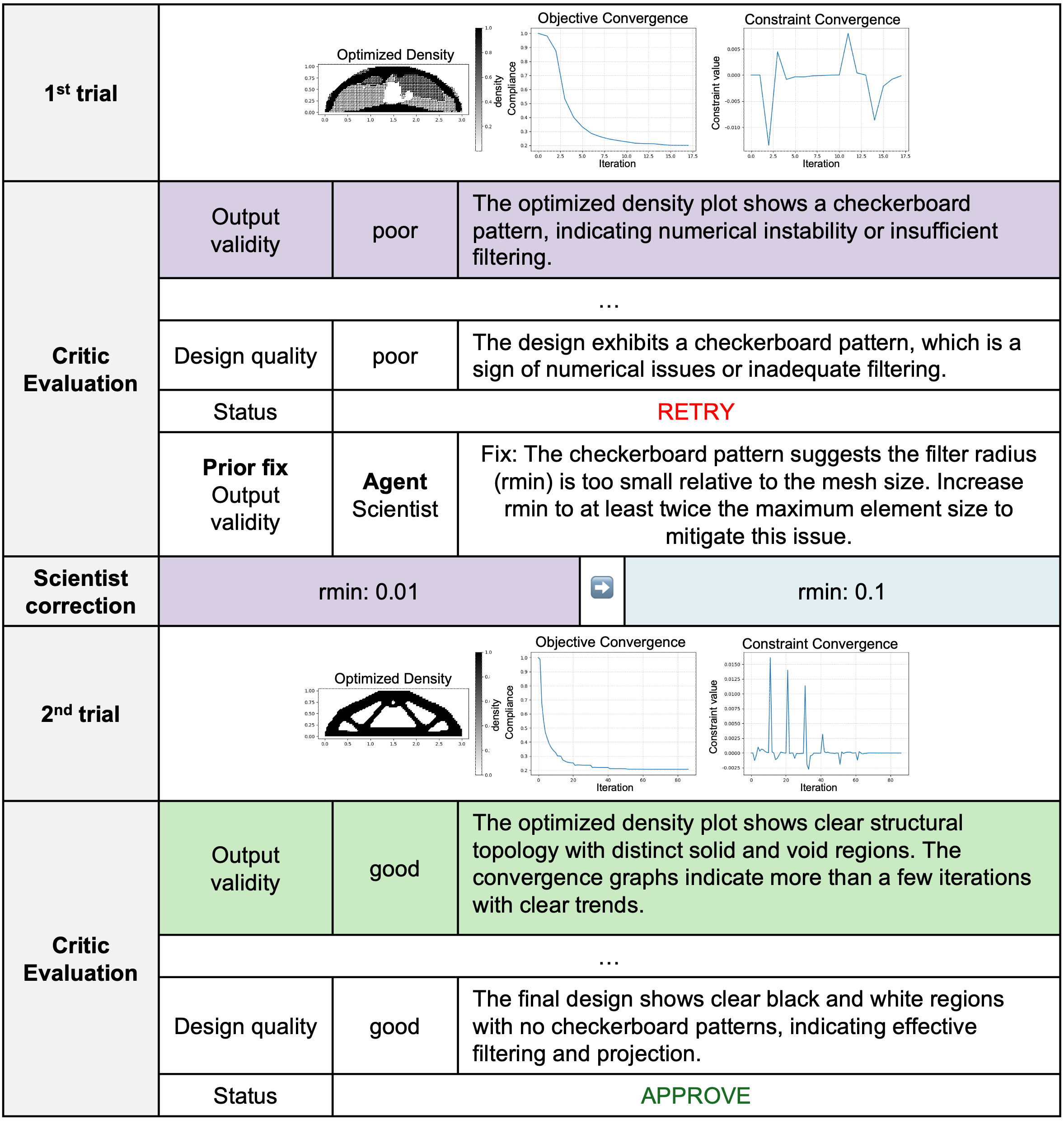}
	\caption{Example of Critic agent refining the filter radius \(r_{\min}\) to alleviate the checkerboard pattern.}
	\label{fig.14}
\end{figure*}

\begin{figure*}
	\centering
	\includegraphics[scale=0.85]{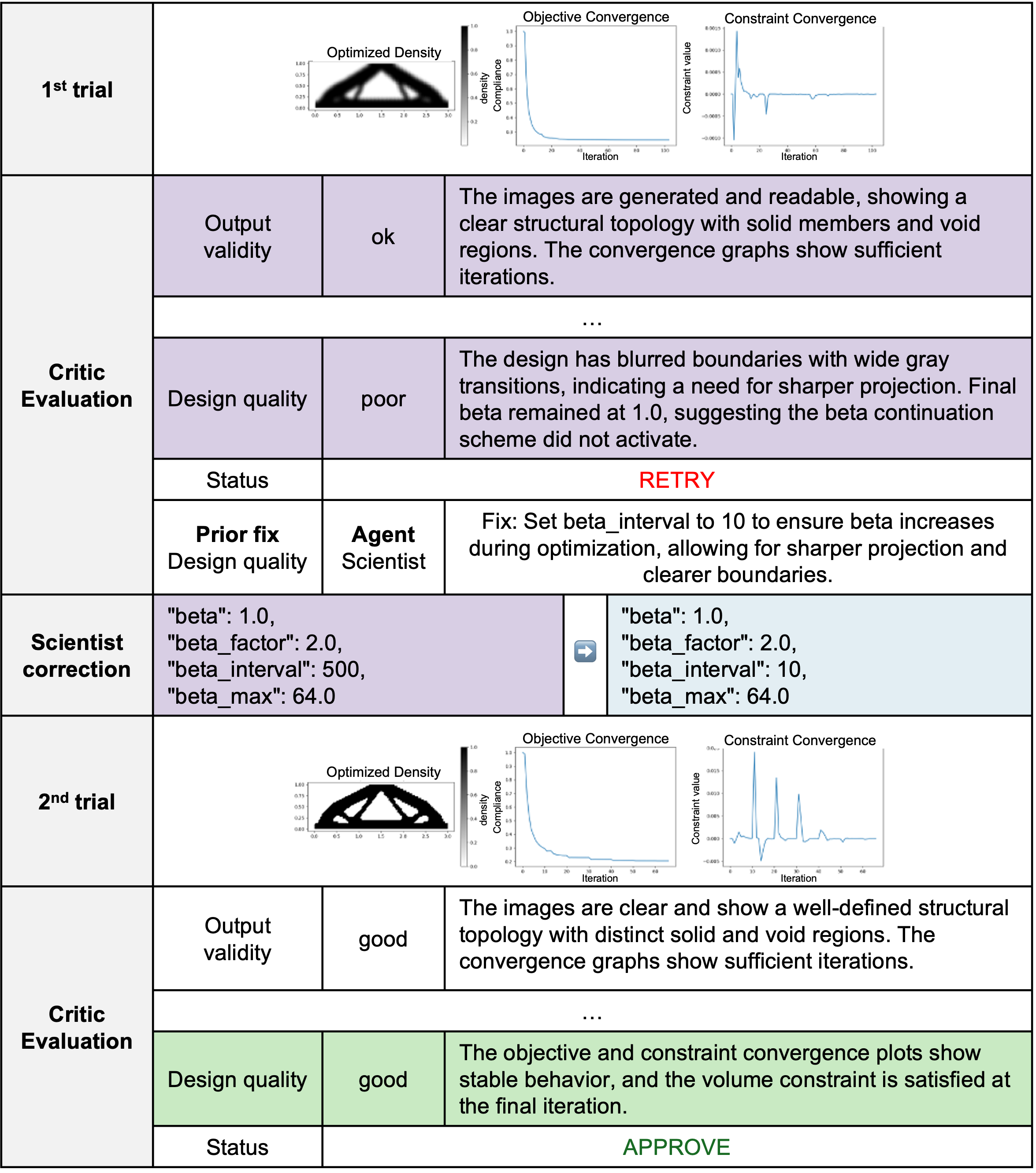}
	\caption{Example of Critic agent refining the beta continuation scheme for a more discrete design.}
	\label{fig.15}
\end{figure*}

Unlike the Validator--Scientist and Reviewer--Coder loops, which intervene before and during code execution respectively, the Critic-driven loop activates after a converged optimization outcome is produced. Failures that pass both prior loops but remain inconsistent with the user query or the standards of acceptable design can be diagnosed only by inspecting the result itself. Critic identifies the unsatisfied criterion, names the agent responsible for refinement, and issues a strategy that re-enters the workflow at the appropriate stage. A single-pass LLM has no analog of this stage: once it produces a converged design, the generation terminates and any inconsistency in the output is not revisited. This role parallels the use of LLMs as evaluators of model outputs, a setting characterized in recent surveys \cite{gu2024survey}. The three figures below illustrate Critic-driven refinement on two failure modes that the upstream loops could not catch: converged designs whose specification turned out to be inconsistent with the user query (Figs.~\ref{fig.13}(a, b)), and converged designs whose layout exhibits known pathologies of topology optimization under inadequate numerical parameters (Figs.~\ref{fig.14}, \ref{fig.15}).

The first mode is illustrated in Fig.~\ref{fig.13}. In Fig.~\ref{fig.13}(a), the constraint function was formulated as the von Mises stress, rather than the volume fraction constraint specified in the user query. The execution converged without runtime errors, so the Reviewer--Coder loop did not flag it, and the Validator--Scientist loop had not caught the formulation inconsistency either. Critic identified the mismatch between the converged outcome and the user's stated constraint, attributed the failing criterion to problem-formulation consistency, and returned the case to Scientist for reformulation. Fig.~\ref{fig.13}(b) shows a boundary-condition variant of the same failure mode: the converged design placed the load in the middle of the right edge, whereas the user query had specified the upper-right end of the arm. In both cases, the failure is invisible to any check that does not compare the converged output to the user's original query.

Figures~\ref{fig.14} and~\ref{fig.15} demonstrate the second mode. In Fig.~\ref{fig.14}, the initial filter radius $r_{\min}$ was set too small relative to the element size, producing a checkerboard pattern in the converged layout. Critic detected this pattern directly from the design visualization, attributed the unsatisfied criterion to design quality, and issued a refinement instruction to increase $r_{\min}$. Fig.~\ref{fig.15} addresses a different aspect of design quality. The projection parameter $\beta$ in the initial continuation schedule grew too gently to drive the density field toward a discrete $0/1$ layout, leaving substantial gray regions in the final design. Identifying these gray regions from the converged layout image, Critic revised the schedule so that $\beta$ increases more aggressively at later iterations, after which the design exhibited the expected discreteness. 
Both cases highlight Critic's reliance on multimodal evaluation, as these failure modes are not legible from the textual convergence log and require interpretation of the design image. We further note that the failures Critic catches at this stage, such as checkerboarding and excessive gray regions, correspond to well-documented numerical instabilities in topology optimization. This suggests that Critic's effectiveness here is currently bounded by what is well represented in the pre-training corpus, and that this scope would broaden as the underlying LLMs improve or are fine-tuned on domain-specific data.

\subsection{User communication} \label{sec:user}

TopOptAgents provides two components for user communication beyond the internal agents. The first is a report generated after every successful execution, which presents the workflow's outcome together with the reasoning that produced it (Sec.~\ref{sec:report}). The second is a mechanism that lets the user submit a comment after the workflow terminates and initiate a further refinement cycle (Sec.~\ref{sec:human_feedback}). A chat interface that displays the agent conversations in real time is used by both.

\subsubsection{Report generation} \label{sec:report}

The report is generated after every successful execution and explains the workflow's outcome to the user. It consolidates the mathematical formulation of the topology optimization problem, the generated Python program, and Critic agent's evaluation of the result. A natural-language narrative in the language of the original user query accompanies these elements and explains them to the user.

\begin{figure*}
	\centering
	\includegraphics[scale=0.9]{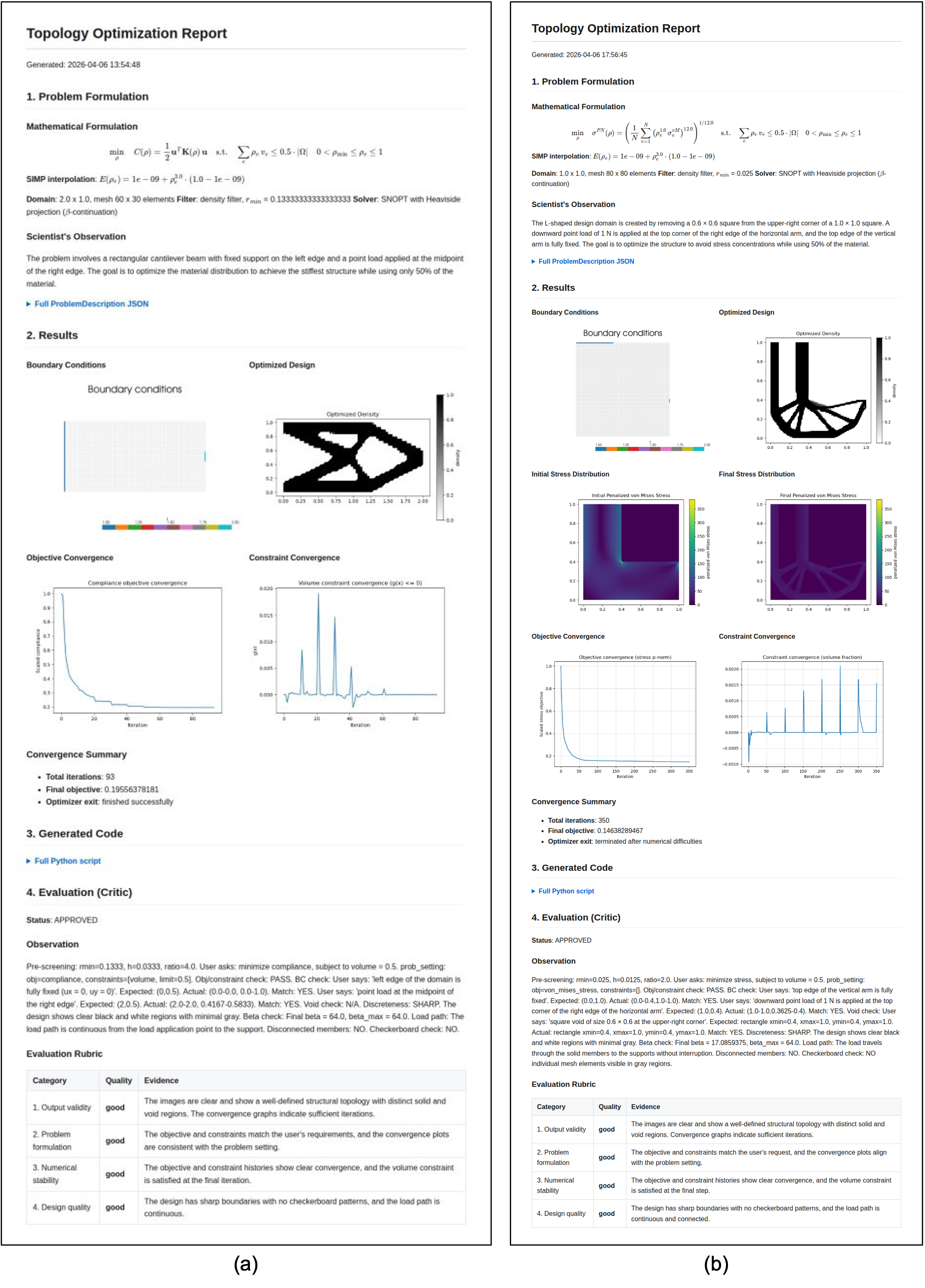}
	\caption{Example of the final report generated by TopOptAgents, including the problem formulation, generated Python code, and result evaluation. (a) Compliance minimization of the cantilever beam, (b) Stress minimization of the L-shaped beam.}
	\label{fig.16}
\end{figure*}

Figure~\ref{fig.16} shows two representative reports, corresponding to the cantilever compliance benchmark in panel (a) and the L-shaped stress benchmark in panel (b). Both panels arrange the same content in the same order, with the problem formulation in mathematical notation appearing first, followed by the executable Python code generated by Coder, and concluded by Critic agent's evaluation referencing the design layout image. The same arrangement handles both compliance and stress formulations without modification, since the report's structure is independent of the specific problem being solved. This multimodal composition, in which mathematical expressions, source code, and design visualization are rendered together, parallels Critic agent's own multimodal evaluation described in Sec.~\ref{sec:critic}. Both rely on the system's ability to carry information across modalities rather than reducing the workflow's output to a single textual stream.

\subsubsection{User-initiated refinement loop} \label{sec:human_feedback}

The chat interface displays the conversations among the agents throughout the multi-agent workflow, allowing the user to monitor each stage as it runs. Once the workflow terminates successfully, the same interface accepts a follow-up comment from the user that requests a structural change to the problem and initiates a further refinement cycle.

\begin{figure*}
	\centering
	\includegraphics[scale=0.5]{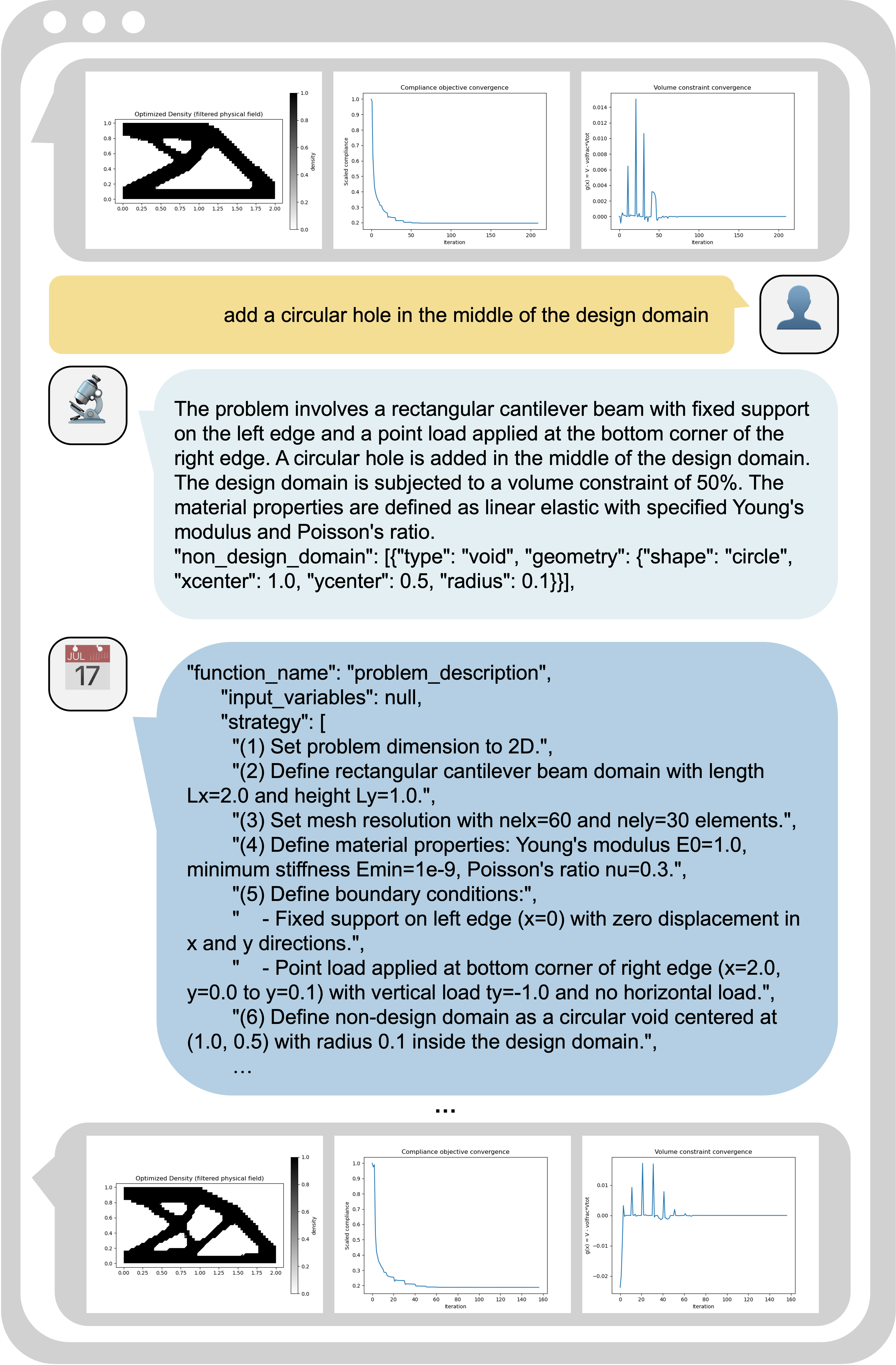}
	\caption{Example of human feedback reflected in a subsequent execution through adding a hole in the problem domain.}
	\label{fig.17}
\end{figure*}

Figure~\ref{fig.17} shows an instance in which the user submits the comment \textit{"add a hole"} to the converged cantilever design. The comment is received by Scientist, which incorporates it into a revised problem specification, and the same agent set re-executes the Validator--Scientist refinement, the Planner stage, the code generation team, and the Critic evaluation on the updated specification. Across the cycle, the agents share the same memory state described in Sec.~\ref{sec:Agent config}, retaining the prior problem formulation, the prior generated code, and Critic's previous verdict, so they adapt to the requested change rather than restart from scratch. The cycle terminates when Critic again approves the new converged design under the same evaluation rubric.

The user-initiated refinement loop is implemented using the agent set and memory state described in Sec.~\ref{sec:Agent config}, so that it does not require any additional infrastructure other than the existing multi-agent workflow. The framework's value, demonstrated through the success-rate recovery in Sec.~\ref{sec:baseline_compare}, therefore extends from automated correction of LLM-internal errors to revision of the design objective in response to user input.

\section{Conclusion} \label{sec:conclusion}

This study presented TopOptAgents, a multi-agent framework that performs topology optimization autonomously, with agents specialized for problem formulation, planning, code generation, execution, and result evaluation interacting through iterative self-refinement loops. The configuration is motivated by the requirements of autonomous topology optimization, which include not only the generation of an initial design but also, like a domain expert's manual workflow, the iterative assessment of the converged outcome and, when needed, the revision of the problem specification, the numerical parameters, or the generated code based on that assessment. To support this assessment-and-revision cycle, the framework introduces three internal refinement loops, each operating at a distinct stage of the pipeline. The Validator--Scientist loop handles specification-level failures before any code is written, the Reviewer--Coder loop handles execution-stage failures from runtime signals, and the Critic-driven loop addresses converged outcomes that pass all prior loops but remain inconsistent with the user query or with acceptable design quality. A fourth loop, triggered after the workflow terminates, allows the user to submit a follow-up comment and runs through the same agent set and shared memory state. Across all loops, the framework bases its output on established components of topology optimization, including problem formulation and sensitivity analysis, so that the generated designs remain consistent with standard numerical procedures. The user accesses the workflow through a chat interface that displays the agent conversations in real time and receives a natural-language report at the end of each run. Evaluated on three benchmark problems selected to vary in their prevalence in the pre-trained LLM's training distribution, TopOptAgents recovered the success rate on problems where the single-pass baseline LLM's performance degraded, with the gap widening on problems further from those the baseline solves successfully in a single pass.

The current framework relies on the pre-training of the underlying LLMs both for recognizing design-quality issues that are widely documented in the topology optimization literature, such as checkerboard patterns and insufficient discreteness, and for specifying common problem formulations such as compliance minimization. Less commonly documented formulations, such as stress minimization with constraint aggregation, are handled in the present study through conditionally adding relevant references to the runtime context, which supplies the agents with the relevant references without modifying the underlying models. Failure modes and formulations that are neither extensively covered nor available as supplied context remain outside the framework's current reliable scope. The most direct approach to expanding this scope is fine-tuning the underlying language models on topology optimization data, which would broaden the set of formulations Scientist agent can specify and the set of design-quality issues Critic agent can recognize \cite{prabhakar2025omniscience}. A complementary approach is retrieval-augmented generation grounded in a HyperGraph ontology of optimization knowledge \cite{stewart2026graphagents, luo2025hypergraphrag}, which would retrieve formal references at runtime for problem formulations with limited coverage in existing tutorials and examples and make them available to agents without modifying the underlying models. Further extensions, including three-dimensional topology optimization and integration with manufacturing process parameters, remain open directions but are not directly addressed by the present results.

Taken together, the present results indicate that a multi-agent self-refinement framework recovers the success rate of LLM-driven topology optimization on problems where a single-pass baseline model fails to converge to a valid design, and that the same agent set accommodates user-driven revision of the design objective once the workflow terminates. Broadening the framework to various formulations through model adaptation or retrieval-augmented reasoning is the direction most directly indicated by the limitations observed in this study.

\section*{CRediT authorship contribution statement}
\textbf{Hyunjee Park}\: Conceptualization, Methodology, Software, Investigation, Validation, Data curation, Formal analysis, Visualization, Writing - original draft. \textbf{Hayoung Chung}\: Conceptualization, Writing - original draft, Writing - review \& editing, Resources, Supervision, Project administration, Funding acquisition.
\section*{Declaration of competing interest}
 The authors declare that they have no known competing financial interests or personal relationships that could have appeared to influence the work reported in this paper.
\section*{Acknowledgment}
This research was supported by the National R\&D Program through the National Research Foundation of Korea (NRF), funded by the Ministry of Science and ICT (No. RS-2025-23525252), and by the InnoCORE program of the Ministry of Science and ICT (No. N10250154). This work was also supported by the Korea Institute of Energy Technology Evaluation and Planning (KETEP) and the Ministry of Trade, Industry \& Energy (MOTIE) of the Republic of Korea (No. RS-2023-00240918).
\appendix
\section*{Data availability}
Data will be made available on request.

\bibliography{main}

\end{document}